\documentclass[10pt,journal,a4paper,final,oneside,twocolumn]{IEEEtran}
\usepackage[square,sort&compress,numbers]{natbib}
\usepackage{times}
\usepackage{graphicx}
\usepackage{cite}
\usepackage{color}
\usepackage{psfrag}
\usepackage{subfigure}
\usepackage{amssymb}
\usepackage{multirow}
\usepackage{epsfig}
\usepackage{pifont}
\usepackage{amsmath}
\usepackage{array}
\usepackage{multicol}
\usepackage[T1]{fontenc}
\usepackage{comment}
\usepackage{acronym}
\usepackage{url}
\usepackage{bm}
\usepackage{cuted}
\usepackage{flushend}
\usepackage{enumerate}
\usepackage{wrapfig}
\usepackage{algorithm}
\usepackage{algpseudocode}
\usepackage{amsmath,amssymb,amsfonts}

\makeatletter
\def\ifundefined{\@ifundefined}
\makeatother 
\begin{document}
\title{{Localization in Reconfigurable Intelligent Surface Aided mmWave Systems: A Multiple Measurement Vector Based Channel Estimation Method}
\author{Kunlun~Li, Jiguang He, {\em Senior Member, IEEE}, Mohammed~El-Hajjar, {\em Senior Member, IEEE} and \\Lie-Liang~Yang, {\em Fellow, IEEE}}
\thanks{K. Liu, M. EI-Hajjar and L.-L. Yang are with the School of Electronics and Computer Science, University of Southampton, SO17 1BJ, UK. (E-mail: kl2u17,meh,  lly@ecs.soton.ac.uk).}
\thanks{J. He is with the Technology Innovation Institute, Abu Dhabi, United Arab Emirates, and also with Centre for Wireless Communications, University of Oulu, Oulu 90014, Finland (e-mail: jiguang.he@tii.ae).}}

\maketitle
\begin{abstract}
The sparsity of millimeter wave (mmWave) channels in the angular and temporal domains is beneficial to channel estimation, while the associated channel parameters can be utilized for localization. However, line-of-sight (LoS) blockage poses a significant challenge on the localization in mmWave systems, potentially leading to substantial positioning errors. A promising solution is to employ reconfigurable intelligent surface (RIS) to generate the virtual line-of-sight (VLoS) paths to aid localization. Consequently, wireless localization in the RIS-assisted mmWave systems has become the essential research issue. In this paper, a multiple measurement vector (MMV) model is constructed and a two-stage channel estimation based localization scheme is proposed. During the first stage, by exploiting the beamspace sparsity and employing a random RIS phase shift matrix, the channel parameters are estimated, based on which the precoder at base station and combiner at user equipment (UE) are designed. Then, in the second stage, based on the designed precoding and combining matrices, the optimal phase shift matrix for RIS is designed using the proposed modified temporally correlated multiple sparse Bayesian learning (TMSBL) algorithm. Afterwards, the channel parameters, such as angle of reflection, time-of-arrival, etc., embedding location information are estimated for finally deriving the location of UE. We demonstrate the achievable performance of the proposed algorithm and compare it with the state-of-the-art algorithms. Our studies show that the proposed localization scheme is capable of achieving centimeter level localization accuracy, when LoS path is blocked. Furthermore, the proposed algorithm has a low computational complexity and outperforms the legacy algorithms in different perspectives.
\end{abstract}


\section{Introduction}\label{introduction.sec}
\subsection{Motivation and Background}
Recently, there has been a surge of interest in localization due to its practical application in various fields, such as intelligent transportation systems and unmanned aerial vehicles \citep{del2018whitepaper}. Traditional localization methods like the global positioning system (GPS) have limited accuracy and high latency, particularly in indoor environments \citep{8226757}. Due to this, the use of millimeter wave (mmWave) based localization techniques has gained significant attention owing to its merits of allowing to employ a large number of antenna elements to achieve high beamspace resolution \citep{8207426}. However, mmWave also faces some practical challenges, including the high path-loss, which can be mitigated by the beamforming provided by massive antenna arrays \citep{6834753,8207426}, and the possible line-of-sight (LoS) blockage, which is suggested to be mitigated by the employment of reconfigurable intelligent surface (RIS) \citep{wirelessthroughreconfigurable}.

In wireless research communities, RISs have recently gained significant emphasis for their ability to control the signal propagation environments for performance improvement \citep{wirelessthroughreconfigurable,zhao2019survey,chen2021towards}. A RIS is composed of numerous reflectors positioned between a transmitter and a receiver, where each individual element of RIS can alter the phase and/or amplitude of the impinged signal \citep{pan2021overview}. This enables the improvement of communication's energy efficiency, spectrum efficiency, positioning accuracy, communication security, etc. \citep{wu2019towards,wymeersch2020radio}. While estimating the cascaded channels in the RIS-supported systems is a challenging task \citep{wu2019intelligent,10071555,wu2019beamforming}, the principles of passive beamforming or the combination of active and passive beamforming, as outlined in \citep{wu2019intelligent,10071555,wu2019beamforming,zhang2022adaptive,shahmansoori2017position}, can be employed for localization purpose.

Thanks to their high angular and temporal resolution, mmWave signals have a high potential for accurate localization, where the angles can be estimated from the angular sparsity and the delay can be estimated with the aid of, such as, the wideband multi-carrier signaling \citep{zhang2022adaptive,shahmansoori2017position,koirala2019localization,lin20183,wang2018pursuance,9493736,zhou2019successive}. In \citep{zhang2022adaptive}, a multi-stage codebook-based adaptive channel estimation algorithm was presented, which enables angular estimation while striking a beneficial trade-off between performance and implementation complexity. In \citep{9493736}, the LoS path was firstly employed to determine the user equipment's (UE) location. Afterwards, the scatterer's (SC) location is estimated as a vertex of a triangle formed by the SC, the estimated UE and BS. However, when the LoS path is obstructed, the accuracy of non-line-of-sight (NLoS) path-based localization is adversely affected by significant reflection loss \citep{10071555}. For instance, in \citep{shahmansoori2017position}, it was demonstrated that the LoS-dependent localization considerably {outperforms the} NLoS-based localization in the conventional mmWave systems. 

Research has shown that the performance of the LoS-based localization is far superior to that of the NLoS-based localization \citep{shahmansoori2017position}. As a result, when LoS is blocked, the RIS-aided localization may provide a promising alternative for improving the localization accuracy in mmWave systems. In \citep{zhou2019successive}, a successive localization and beamforming strategy was proposed to achieve the joint UE localization and channel estimation in mmWave multi-input-multi-output (MIMO) systems.

In literature, numerous studies have been conducted on the RIS-assisted localization \citep{9729782,win2022location,he2020adaptive,zhang2020towards,ning2021terahertz,10071555}. Specifically in \citep{9729782}, the concept of continuous intelligent surface (CIS) was introduced, and the limits of RIS-aided localization and communication systems were discussed. More specifically, in \citep{9729782}, a general signal model for the RIS-aided localization and communication systems was presented, when considering both the far and near-field scenarios. Both CIS and discrete intelligent surface (DIS) were introduced to enhance the localization accuracy and spectral efficiency through RIS phase response optimization. In \citep{win2022location}, the holographic network localization and navigation (NLN) were explored, where RISs were used in the controlled electromagnetic environments by utilizing the polarization and specific antenna patterns. The study shows that RISs have the potential to improve the robustness of the holographic localization against obstructions. In \citep{he2020adaptive}, a RIS-assisted localization scheme supported by the adaptive beamforming design using the hierarchical codebook algorithm was proposed for the joint localization and communication, when assuming the absence of LoS path. In \citep{zhang2020towards}, a received signal strength (RSS)-based positioning scheme was investigated in a RIS-aided mmWave system. Furthermore, in \citep{ning2021terahertz}, a multi-RIS-assisted multiple-user beam training scheme was proposed for Terahertz systems, where a ternary-tree hierarchical codebook-based algorithm was employed to reduce complexity. Furthermore, in \citep{10071555}, a joint active and passive beamforming codebook based localization scheme was proposed for the RIS-aided mmWave systems, when assuming that the LoS link's state is unknown. It is shown that the scheme is capable of achieving a good trade-off between performance and training overhead.

\subsection{Research Problem and Contributions}
Based on the research background, the authors of \citep{9493736} proposed a sparse Bayesian learning (SBL) algorithm assisted localization scheme to exploit the beamspace channel sparsity, where the single measurement vector (SMV) model was employed for channel estimation. Although the SMV model has the advantages of simplicity and low hardware complexity, the counterpart multiple measurement vector (MMV) model is capable of improving the robustness and positioning accuracy, and also allows to employ the time/frequency/spatial diversity \citep{zhang2011sparse}. Thus, in this paper, we focus on the MMV model with the SMV model being used as the benchmark for performance comparison. In literature, many methods, such as the approximate message passing (AMP), orthogonal-AMP (OAMP) algorithms have been introduced to take the MMV-based problems \citep{kim2011belief,cheng2023orthogonal,shao2019dimension}. However, these algorithms have imposed constraints on the sensing matrix, requiring it satisfy the restricted isometry property (RIP), and follow the near Gaussian distribution to make Gaussian approximation effective \citep{kim2011belief}. Moreover, in \citep{9493736, shahmansoori2017position}, the RIS was not employed, and the localization is achieved via estimating the LoS channel's parameters. However, as above-mentioned, the LoS path in mmWave communications has a high probability of being blocked. In this case, the employment of RIS provides a promising solution for localization. Furthermore, we consider the RIS-assisted MMV model with the multi-carrier mmWave systems, i.e. orthogonal frequency division multiplexing (OFDM) mmWave systems. Comparatively, OFDM was not considered in \citep{he2021channel}, and the time of arrival (ToA) is unable to be estimated. By contrast, in our MMV model, with the aid of the information extracted from both time domain and frequency (subcarriers) domain, both the angle and ToA information can be estimated.

More specifically, in this paper, we design and investigate a localization scheme for the RIS-aided mmWave system with OFDM signaling. Our localization scheme consists of two stages. During the first stage, the angle of departure (AoD) at base station (BS) and the angle of arrival (AoA) at UE for designing the precoding and combining matrices are estimated by exploiting beamspace sparsity. In the second stage, the time and frequency domain observations are employed to build the MMV model for estimating the channel parameters and designing the RIS phase shift matrix. Finally, the localization of UE is obtained from the estimated channel parameters. 
More explicitly, our contributions can be summarized as follows.
\begin{itemize}
    \item A two-stage localization algorithm for the OFDM mmWave systems is proposed to allow a UE to estimate its own location in a single-BS single RIS system, when the LoS path between BS and UE is assumed to be blocked.
    \item The localization is formulated as a compressed sensing (CS) problem, where the channel parameters, such as AoA, AoD, angle of reflection (AoR) and ToA, are relied on the physical locations of BS, UE, SCs and RIS. The localization problem is solved in two stages. In the first stage, the conventional beamspace formulation is introduced to estimate the channel parameters of the channel between BS and UE, when randomly setting the RIS's phase shift matrix, based on which the precoding and combining matrices used by BS and UE are designed. Based on the designed precoding and combining matrices, in the second stage, the observations from both time and frequency domain are utilized to build the MMV model for optimizing the RIS phase shift matrix. Afterwards, the channel parameters of the AoR at RIS and ToA between RIS and UE are estimated to ultimately estimate the UE's location.
    \item A large number of elements deployed at RIS leads to high computational complexity on the cascaded channel estimation and localization, especially for the SBL-based algorithms. Inspired by \citep{zhang2011sparse}, a modified temporally correlated multiple sparse Bayesian learning (TMSBL) algorithm is proposed for the RIS-aided localization systems. The proposed algorithm is able to reduce the size of the matrices involving inversion and hence the implementation complexity.
    \item We compare the localization performance of the proposed modified TMSBL algorithm with the state-of-art algorithms, including the orthogonal matching pursuit (OMP), distributed compressed sensing simultaneous orthogonal matching pursuit (DCS-SOMP), SBL and group SBL (GSBL) algorithms. Our studies and simulation results show that our proposed modified TMSBL algorithm outperforms the OMP, DCS-SOMP and SBL algorithms in terms of the localization accuracy, when the number of training blocks is relatively large. At the same time, it has much lower complexity than the conventional SBL and GSBL algorithms. Furthermore, we demonstrate and analyze the impact of different system parameters, such as the number of training blocks, beamspace resolution and the placement of RIS, on the localization accuracy. 
    
\end{itemize}

\subsection{Organization of the Paper and Notations}

The rest of the paper is organized as follows. Section~\ref{System Model.sec} introduces the system and signaling models for localization. Section~\ref{problemformulationpaper3.sec} introduces the framework of the two stage channel estimation based localization scheme. Section~\ref{proposedSBL.sec} presents the localization problem formulation with the conventional GSBL and the proposed modified TMSBL algorithms. The simulation setup and simulation results are provided in Section \ref{Simulationsetup.sec}. Finally, we concisely conclude this paper in Section \ref{conclusion.sec}.

{\textit{Notations}: $a$, ${\mathbf{a}}$, ${\mathbf{A}}$ stand for scalar, vector and matrix, respectively. ${{\mathbf{A}}^{\rm T}}$, ${\mathbf{A}}^{\rm H}$, ${\mathbf{A}}^\dag$, ${\left\| {\mathbf{a}} \right\|_{2}}$ and ${\left\| {\mathbf{A}} \right\|_{F}}$ represent transpose, Hermitian transpose, pseudoinverse, Euclidean norm and Frobenius norm of matrix ${\mathbf{A}}$, respectively. The (${i,j}$)-th entry of ${\mathbf{A}}$ is $[{\mathbf{A}}]_{i,j}$, and $\text{diag}(\mathbf{a})$ is a diagonal matrix formed by the diagonal elements of $\mathbf{a}$. $\text{Trace}(\bf A)$ denotes the trace of matrix $\bf A$}, $\mathbb{E}(\bf A)$ is the expectation of $\bf A$, $\text{vec}(\bf A)$ is the vectorization operation of $\bf A$, $\text{mod}(i,j)$ denotes the modulo operation, and $j = \sqrt{-1}$. $(\bf A)^*$ denotes the conjugate of matrix $\bf A$.

\section{System Model}\label{System Model.sec}
\begin{figure}[!hbt]
	\centering
	\includegraphics[width=1\linewidth]{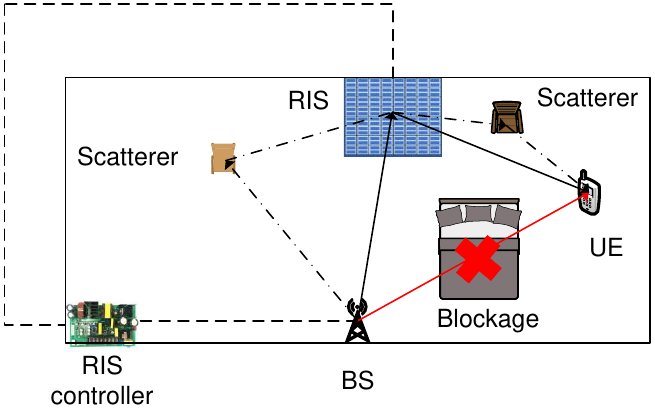} 
	\caption{Illustration of RIS aided localization model.}
	\label{fig:systemmultipleRISs}
\end{figure}

In this paper, a downlink MIMO mmWave localization system with a single BS and a single UE is considered, which are equipped with $N_{\text B}$ and $N_{\text M}$ antennas, respectively, as shown in Fig. \ref{fig:systemmultipleRISs}. A RIS with $N_{\text R}$ antennas is employed to overcome the line-of-sight (LoS) blockage between BS and UE, as shown in Fig. \ref{fig:systemmultipleRISs}. More explicitly, the OFDM modulated downlink signals and the downlink position reference signals (PRSs) are transmitted from BS, which are then reflected by RIS to UE, due to the LoS blockage. Based on the reflected signals received by UE, the UE estimates the channel parameters as well as its location.

Let the location of BS be $\mathbf{b}=[b_x,b_y]^{\text T}\in $ $\mathbb{R}^2$, and that of UE be ${\mathbf{m}} = {[{m_{x}},{m_{y}}]^\text T} \in {\mathbb{R}^2}$, while the location of RIS is given as ${\mathbf{r}} = {[{r_{x}},{r_{y}}]^\text T} \in {\mathbb{R}^2}$. Furthermore, one scatterer is assumed between BS and RIS, and another scatterer exists between RIS and UE, whose locations are expressed as ${\mathbf{s}_l} = {[{s_{x,l}},{s_{y,l}}]^\text T} \in {\mathbb{R}^2},l=1,2$. The locations of BS and RIS are fixed and assumed to be known, while the location of UE is to be estimated. As shown in Fig.~\ref{fig:systemmultipleRISs}, the LoS path between BS and UE is blocked. In this case, the RIS is employed to overcome the blockage problem and improve the localization coverage. Using the received signal reflected by RIS, UE can estimate the channel from RIS. Then, based on the channel estimation results and the \textit{a~prior} location information (i.e, $\mathbf{b}$ and $\mathbf{r}$) of BS and RIS, UE can estimate its location. Below we present the transmitter processing, channel modeling and the receiver processing.
\subsection{Transmitter Model}
The aim of design is for the UE to estimate its location based on the PRS transmitted from BS and reflected by RIS, when OFDM signalling is employed. In detail, at BS transmitter, $N_{\text D}$ data streams are first precoded by the digital beamformer {${\bf F}_{\text {BB}}[n]\in \mathbb{C}^{N_{\rm{RF}} \times N_{\text D}}$} for the $n$-th subcarrier, where $N_{\rm{RF}}$ denotes the number of radio frequency (RF) chains. Then, for each RF chain, a $N$-point inverse fast Fourier transform (IFFT) is implemented to transform the data from frequency-domain to time-domain. Afterwards, a cyclic prefix (CP) is added before the RF level precoding. We assume that the CP length of $T_{\rm{CP}} = S_{\rm{CP}}T_{\rm{s}}$ is longer than the channel's delay spread, where $T_{\rm{s}}$ is the sampling interval and ${S_{{\rm{CP}}}}$ denotes the CP length in samples. Following the CP insertion, an analog beamformer ${\bf F_{\text {RF}}}\in \mathbb{C}^{{N_\text{B}} \times{N_{{\rm{RF}}}} }$ is implemented, which is independent of the subcarriers. Overall, the hybrid beamformer is expressed as ${\bf F}[n]={\bf F}_{\text {RF}}{\bf F}_{\text {BB}}[n]$, which satisfies ${\left\| {{{\bf{F}}_{{\rm{RF}}}}{{\bf{F}}_{{\rm{BB}}}}[n]} \right\|_F} = 1$. 

Let the $g$-th OFDM symbol transmitted for positioning be expressed as ${{\bf{x}}^g}[n] = {[{x_1}[n],{x_2}[n],...,{x_{{N_D}}}[n]]^\text{T}} \in {\mathbb{C}^{{N_D}}}$,  $n=1,...,N$; $g=1,...,G$. Then, the transmitted position reference signals can be expressed in baseband form as: 
\begin{align}\label{transmittedPRS.eq}
{{\mathbf{\vec{x}}}^g[n]} = \sqrt P {{\mathbf{F}}^g}[n]{{\mathbf{x}}^g}[n]\in\mathbb{C}^{N_\text{B}\times 1},
\end{align}
where $P$ is the transmit energy, ${{\mathbf{F}}^g}[n]$ is the hybrid directional beamforming matrix for the $g$-th reference signal, which satisfies ${\text{Trace(}}{{\mathbf{F}^g[n]}^{\text{H}}}{\mathbf{F}^g[n]}{\text{) = 1}}$ and $ 
\mathbb{E}\{ {\mathbf{x}}^g[n]{{\mathbf{x}}^g}[n]^{\text{H}}\}  = {{\mathbf{I}}_{{N_D}}}$. 

\subsection{Channel Model}
Consider a mmWave channel, whose parameters, including AoA, AoD/AoR, and ToA, are determined by the locations of BS, UE and RIS. We assume that when BS is communicating with UE, the LoS path between them is blocked. Therefore, the downlink transmission has to rely on the virtual line-of-sight path (VLoS), as depicted in Fig.~\ref{fig:systemmultipleRISs}. The AoD of BS and the AoR of RIS are termed as $\theta_{\text {BR}}$ and $\theta_{\text {RM}}$, while the AoA of RIS and that of UE are expressed as $\phi_{\text {BR}}$ and $\phi_{\text {RM}}$, respectively. Moreover, the ToA between BS and RIS, and that between RIS and UE are denoted as $\tau_{\text {BR}}$ and $\tau_{\text {RM}}$, respectively. The study assumes uniform linear array (ULA). Hence, the steering vectors $\mathbf{a}_{\text{B},n}(\theta_{\text {BR}})$ and $\mathbf{a}_{\text{R},n}(\phi_{\text {BR}})$ at BS and RIS in the context of subcarrier $n$ are \citep{9200524}
\begin{align}
   \mathbf{a}_{\text{B},n}(\theta_{\text {BR}})=&\frac{1}{\sqrt{N_\text{B}}}
\left[1,{e^{-j2\pi\frac{d}{\lambda_n}\sin(\theta_{\text {BR}})}},...,\right.\nonumber\\
&\left.e^{-j2\pi\frac{d}{\lambda_n}\sin(\theta_{\text {BR}})(N_{\text{B}}-1)}\right]^\text{T}\in\mathbb{C}^{N_{\text{B}}\times 1}, \\
   \mathbf{a}_{\text{R},n}(\phi_{\text {BR}})=&\frac{1}{\sqrt{N_\text{R}}}
\left[1,{e^{-j2\pi\frac{d}{\lambda_n}\sin(\phi_{\text {BR}})}},...,\right.\nonumber\\
&\left.e^{-j2\pi\frac{d}{\lambda_n}\sin(\phi_{\text {BR}})(N_{\text{R}}-1)}\right]^\text{T}\in\mathbb{C}^{N_{\text{R}}\times 1},  
\end{align}
where $d$ represents the antenna spacing between two adjacent elements, and $\lambda_n$ is the wavelength of the $n$-th subcarrier. For simplicity, we assume that the signal bandwidth $B \ll {f_c}$, yielding $\lambda_n \approx \lambda_c$, where $\lambda_c$ represents the wavelength of the main carrier \citep{shahmansoori2017position,7400949}. Then, the mmWave channel model introduced in \citep{chen2021beam,chen2020hybrid} can be applied to obtain the ($N_{\text R}\times N_{\text B}$)-dimensional frequency domain channel matrix from the BS to the RIS, which can be represented as \citep{9493736}:
\begin{align}\label{channelBS-RIS.eq}
{\mathbf{H}}_{\text{BR}}[n] = {{\mathbf{A}}_\text{R}}({\bf{\phi}_{\text {BR}}})[n]{\mathbf{\Sigma_{\text{BR}} }}[n]{\mathbf{A}}_\text{B}^\text{H}({\bf{\theta}_{\text {BR}}})[n],  
\end{align}
where
\begin{align}\label{steering.eq}
   &\mathbf{A}_{\text{B}}({\bm{\theta}_{\text {BR}}})[n]=[\mathbf{a}_{\text {B},n}(\theta_{\text {BR},0}),\mathbf{a}_{\text {B},n}(\theta_{\text{BR},1})...,\mathbf{a}_{\text{B},n}(\theta_{\text{BR},L_{\text{BR}}-1})],\\
    &\mathbf{A}_{\text{R}}({\bm{\phi}_{\text {BR}}})[n]=[\mathbf{a}_{\text{R},n}(\phi_{\text{BR},0}),\mathbf{a}_{\text{R},n}(\phi_{\text{BR},1})...,\mathbf{a}_{\text{R},n}(\phi_{\text{BR},L_{\text{BR}}-1})], 
\end{align}
and the diagonal matrix ${\mathbf{\Sigma }}_{\text{BR}}[n]$ is given by \citep{10071555}
\begin{align}
\label{diagnal.eq}
{\mathbf{\Sigma }}_{\text{BR}}[n] &= \sqrt {{N_\text{B}}{N_\text{R}}} \nonumber\\
& \times \text{diag}\{ {{\beta _{\text{BR},0}}} {{\rho _{\text{BR},0}}} {e^{\frac{{ - j2\pi n{\tau _{\text{BR},0}}}}{N{T_s}}}},...,\nonumber\\&{\beta _{\text{BR},L_{\text{BR}}-1}}{ {{\rho _{\text{BR},L_{\text{BR}}-1}}} }{e^{\frac{{ - j2\pi n{\tau _{\text{BR},L_{\text{BR}}-1}}}}{N{T_s}}}} \}.
\end{align}
In \eqref{steering.eq}-(\ref{diagnal.eq}), $L_{\rm{BR}}$ is the number of paths between BS and RIS, $\beta _{\text{BR},l}$ and $\rho _{\text{BR},l}$ \footnote{Note that, in mmWave systems, the NLoS paths reflected by SCs experience additional reflection loss, as the result that part of the energy is reflected instead of penetrating or is absorbed when it encounters the surface of an object \citep{al2017simplified}.} are respectively the Rician complex fading gain and path-loss of the $l$-th path between BS and RIS, and $T_S=1/B$ is the sampling period. In \eqref{diagnal.eq}, the time delay $\tau _{\text{BR},l}$, i.e. ToA, is given by $\tau _{\text{BR},l}=d_{\text{BR},l}/c$, where $c$ denotes the speed of light, while $d_{\text{BR},l}$ is the propagation distance of the $l$-th path. Specifically, for LoS path ($l=0$), the distance between BS and RIS is evaluated as $
{d_{\text{BR},0}} = {\left\| {{{\mathbf{r}}} - {{\mathbf{b}}}} \right\|_2}$, while for NLoS path ($l>0$), the distance is 
${d_{\text{BR},l}} = {\left\| {{{\mathbf{s}}_l} - {{\mathbf{b}}}} \right\|_2} + {\left\| {{{\mathbf{r}}} - {{\mathbf{s}}_l}} \right\|_2}$, where ${{\mathbf{s}}_l}$ is the location of SC. 

Similarly, the channel between RIS and UE can be represented as
\begin{align}\label{channelRIS-MS.eq}
{\mathbf{H}}_{\text{RM}}[n] = {{\mathbf{A}}_\text{M}}({\bm{\phi}_{\text {RM}}})[n]{\mathbf{\Sigma_{\text{RM}} }}[n]{\mathbf{A}}_\text{R}^\text{H}({\bm{\theta}_{\text {RM}}})[n], 
\end{align}
where ${{\mathbf{A}}_\text{M}}({\bm{\phi}_{\text {RM}}})[n]$ and ${\mathbf{A}}_\text{R}({\bm{\theta}_{\text {RM}}})[n]$ are defined similarly as that in (\ref{channelBS-RIS.eq}). 

By combining (\ref{channelBS-RIS.eq}) and (\ref{channelRIS-MS.eq}), the frequency domain cascaded channel from BS to UE can be represented as 
\begin{align}\label{cascadedchannel-kthRIS.eq}
{\mathbf{H}}^t[n] =& {\mathbf{H}}_{\text{RM}}[n]{\bf{\Omega}}_t{\mathbf{H}}_{\text{BR}}[n],
\end{align}
where ${\bf \Omega}_t \in\mathbb{C}^{N_{\text{R}}\times N_{\text{R}}}$ is RIS's phase shift matrix for $t=0,\dots,J$, which is composed of the discrete phase shifters. Thus, ${\bf \Omega}_t$ is a diagonal matrix  that is unit-modulus on the diagonal elements \citep{9737373,10025848}. Specifically, the diagonal element $[{\bf \Omega}_t]_{i,i}=\eta e^{j\omega_{i}}$, where $\omega_{i} \in [0,2\pi]$ and $\eta$ denotes the reflection coefficient. Note that, for simplicity and also without any loss of generality, as long as the reflection coefficient is known, we assume that RIS is of perfect reflection, meaning $\eta=1$. More explicitly, if there is reflection loss (imperfect reflection), the ToA cannot be estimated accurately, as the result that the ToA is related to the channel gain. In literature, the perfect reflection is assumed in many RIS-aided channel estimation and localization \citep{he2020adaptive,he2020large,9729782}. 

On the other hand, considering that the RIS phase shift matrix is only related to the angular parameters, i.e., $\theta_{\text {RM},k}$ and $\phi_{\text {BR},k}$ \citep{he2021channel}, according to (\ref{cascadedchannel-kthRIS.eq}), the effective channel can be further expressed as 
\begin{align}\label{effectivechannel.eq}
{\bf H}_{\text{eff}}^t[n]=&{\text{diag}}({\widehat{\bm \rho}}_{\rm{RM}}[n]){\bf A}_{\text R}^{\text H}({\bm \theta}_{\text{RM}})[n]{\bm \Omega}_t{{\mathbf{A}}_\text{R}}({\bm{\phi}_{\text {BR}}})[n]\nonumber\\&\times{\text{diag}}({\widehat{\bm \rho}}_{\rm{BR}}[n]), 
\end{align}
where ${\widehat{\bm \rho}}_{\rm{RM}}[n]=[\beta _{\text{RM},0}\rho _{\text{RM},0}e^{\frac{ - j2\pi n{\tau _{\text{RM},0}}}{N{T_s}}},\dots,{{\beta _{\text{RM},L_{\text{RM}}-1}}}\\{\rho _{\text{RM},L_{\text{RM}}-1}}e^{\frac{ - j2\pi n{\tau _{\text{RM},L_{\text{RM}}-1}}}N{T_s}}]^{\text{T}}$, and ${\widehat{\bm \rho}}_{\rm{BR}}[n]$ is similarly defined according to \eqref{diagnal.eq}. Thus, the frequency domain channel can then be represented as
\begin{align}\label{cascadedchannel2.eq}
{\bf H}^t[n] = {{\mathbf{A}}_\text{M}}({\bm{\phi}_{\text {RM}}})[n]{\bf H}_{\text{eff}}^t[n]{\mathbf{A}}_\text{B}^\text{H}({\bm{\theta}_{\text {BR}}})[n],
\end{align}
where ${\mathbf{H}}_{\text{eff}}^t[n]$ defined in \eqref{effectivechannel.eq} is a function of the phase shifter matrix ${\bf{\Omega}}_t$ for the $t$-th data block, which shows the importance of the angular parameters ${\bm \theta}_{\text {RM}}$ and ${\bm\phi}_{\text {BR}}$, as well as of the corresponding phase shifter design to the channel estimation and localization.

\subsection{Receiver Model}\label{receivermodelpaper3.subsec}
The signals received at UE are used for localization. After the transmission over the mmWave channel as above described, CP removal and FFT processing, the hybrid combining matrix  ${\bf{W}}[n]={\bf W}_{\text {RF}}{\bf W}_{\text {BB}}[n]$ is employed for processing the reflected signals, giving the output denoted as
\begin{align}\label{k-threceivermodel.eq}
    y^g[n] = \sqrt{P}({\bf{w}}^g[n])^{\text H}{\mathbf{H}}^t[n]{{\mathbf{\vec{x}}}^g[n]}+({\bf{w}}^g[n])^{\text H}{\bf n}[n], 
\end{align}
where ${\bf{w}}^g[n]$ is the combining training vector for the $g$-th OFDM symbol, which is the $g$-th column of the combining matrix ${\bf{W}}[n]$. ${\bf n}[n]$ is the corresponding additive white Gaussian noise (AWGN) vector distributed with zero mean and a variance $\sigma^2$ per element.

To sense the location of UE, in this paper, we exploit the sparsity of mmWave channel and propose a two-stage channel estimation scheme, to estimate the channel parameters, including $\bm{\theta} _{\text{RM}}$ and $\bm{\tau} _{{\text{RM}}}$. Note that, as all the location parameters are embedded in the channel impulse responses (CIR) of the cascaded mmWave channel, the localization problem can be converted to a two-stage estimation/design problem, which contains the channel parameter estimation and the phase shifter design, as detailed in the next section.

\section{Problem Formulation}\label{problemformulationpaper3.sec}
\begin{figure}[!hbt]
	\centering
	\includegraphics[width=1\linewidth]{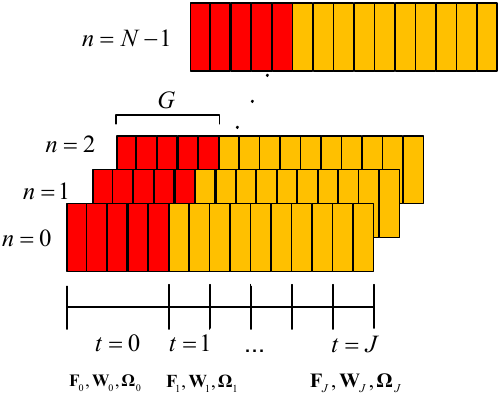} 
	\caption{Illustration of multiple time blocks/subcarriers based channel estimation. The red columns denote the OFDM symbols in the first stage, where $G\gg L_{\text{BR}}$ or $ L_{\text{RM}}$, while the yellow columns denote the OFDM symbols used in the second stage, where the number of columns is equal to the number of propagation paths.}
	\label{fig:CEdesign}
\end{figure}
In the state-of-the-art of the RIS-aided localization, the channel parameters, such as AoR/AoD and ToA, are normally estimated by exhaustively searching the predefined precoding/combining matrices, for the optimal pair of precoding and combining matrices \citep{9729782,wang2021joint,he2020adaptive}. In this paper, we propose a multiple-block based MMV model, based on which we solve the localization problem as a CS channel estimation problem. In this section, the MMV CS model is first formulated. To estimate the reflected angular parameter ${\bm\theta} _{\text{RM}}$ and the time delay parameter ${\bm\tau} _{\text{RM}}$, which are challenging tasks, we separate the localization procedure into two stages, which are termed as the channel estimation and localization stages, as shown in Fig.~\ref{fig:CEdesign}. As shown in Fig.~\ref{fig:CEdesign}, the channel estimation frame is divided into $J+1$ time blocks, and the $N$ subcarriers are employed in all time blocks. More explicitly, for different blocks $t=0,1,...,J$, different phase shift matrices ${\bf{\Omega}}_{t}$ are used by the RIS, meaning that ${\bf{\Omega}}_0 \neq {\bf{\Omega}}_1 \neq ... \neq ... {\bf{\Omega}}_J$. This can be achieved by employing the n-type field-effect transistor (nFET) switches, which guarantee that the RIS phase shift's change frequency is much lower than the symbol rate within one coherence time of mmWave channel \citep{he2021channel}. As above-mentioned, the whole localization process consists of two stages. Explicitly, during the first stage, the conventional CS-aided channel estimation is first employed to estimate the AoD and the AoA, respectively, at BS and at UE. Then, the precoding and combining matrices are designed based on the estimated AoD and AoA, associated with a randomly set RIS phase shift matrix. Note that, there are many CS algorithms, such as, the basis pursuit (BP), message passing (MP), SBL algorithms, which can be employed for the AoD and AoA estimation. As an example, the DCS-SOMP algorithm is employed in this paper for the sake of simplicity. Then, based on the beamformer and combiner designed in the first stage, the MMV CS model is formulated. Then, during the second stage, based on the designed precoding and combining matrices, the AoR and ToA are estimated, based on which the location of UE is estimated. Let us below analyze them in detail.

\subsection{Stage 1: DCS-SOMP Assisted Design of BS Precoder and UE Combiner}
During the first time block corresponding to $t=0$, as shown in Fig.~\ref{fig:CEdesign}, a random precoding matrix ${\bf{F}}^0[n] = [{{\mathbf{\vec{x}}}^1[n]},...,{{\mathbf{\vec{x}}}^{G}[n]}] \in \mathbb{C}^{N_{\text B}\times G }$ with Gaussian distribution and a random combining matrix ${\bf{W}}^0[n] \in \mathbb{C}^{N_{\text M}\times G }$ with Gaussian distribution are employed, where $G$ denotes the number of symbol durations, represented by the red columns in Fig.~\ref{fig:CEdesign}. Note that, $G$ is required to be larger than the number of propagation paths to obtain angular sparsity. Furthermore, in this stage, the phase shift matrix ${\bf{\Omega}}_0$ at RIS is set randomly and fixed during Stage 1. Thus, for the first stage in Fig.~\ref{fig:CEdesign}, the observations corresponding to (\ref{k-threceivermodel.eq}) can be further written as 
\begin{align}\label{simplifiedreceivermodel.eq}
        {\mathbf{Y}}^0[n] = \sqrt{P}({\bf{W}}^0[n])^{\text H}{\mathbf{H}}^0[n]{{\mathbf{F}}^0[n]}+({\bf{W}}^0[n])^{\text H}{\bf N}^0[n].
\end{align}
 Let us introduce the beamspace channel representation \citep{7400949}, which is obtained via uniformly sampling the spatial angles in the beamspace, yielding:
\begin{align}\label{beamspaceRIS1.eq}
{{\bf{U}}_\text B} =& \left[ {{{\bf{u}}_\text B}\left( {{q_0}} \right), \ldots ,{{\bf{u}}_\text B}\left( {{q_{{N_\text B} - 1}}} \right)} \right],\nonumber\\
{{\bf{u}}_\text B}\left( {{q_l}} \right) =& \left[ {{e^{ - j2\pi \frac{{{N_\text B} - 1}}{2}{q_l}}}, \ldots ,{e^{j2\pi \frac{{{N_\text B} - 1}}{2}{q_l}}}} \right]^{\text T}.
\end{align}
 In \eqref{beamspaceRIS1.eq}, ${{\mathbf{U}}_{\text{B}}}$ is a unitary Discrete Fourier Transform (DFT) matrix determined by the beamspace grid indices of {${q_l} =  - \frac{{{N_\text{B}} - 1}}{{2{N_\text{B}}}} + \frac{l}{{{N_\text{B}}}},\forall l \in \left[ {0,{N_\text{B}} - 1} \right]$}. 
Then, the beamspace channel representation of ${\mathbf{H}}^0[n]$ in \eqref{simplifiedreceivermodel.eq} for the $t=0$ block can be written as
\begin{align}\label{beamspaceRIS2.eq}
{{\mathbf{H}}_{\text{v}}^0}[n] = {\mathbf{U}}_{\text{M}}^{\text{H}}{\mathbf{H}}^0[n]{{\mathbf{U}}_{\text{B}}} \in {\mathbb{C}^{{N_{\text{M}}} \times {N_{\text{B}}}}},
\end{align}
where ${{\mathbf{U}}_{\text{M}}}$ for the receiver array is obtained similarly as ${{\mathbf{U}}_{\text{B}}}$. Since in mmWave communications, usually only a small number of propagation paths from transmitter to receiver, the beamspace channel matrix ${{\mathbf{H}}_{\text{v}}}[n]$ is a sparse matrix \citep{6979963}. Upon applying (\ref{beamspaceRIS2.eq}) into (\ref{simplifiedreceivermodel.eq}), we obtain
\begin{align}\label{beamspaceRIS3.eq}
     {\mathbf{Y}}^0[n] = &\sqrt{P}({\bf{W}}^0[n])^{\text H}{\mathbf{U}}_{\text{M}}{{\mathbf{H}}_{\text{v}}^0}[n]{\mathbf{U}}_{\text{B}}^{\text H}{{\mathbf{F}}^0[n]}\nonumber\\
     &+({\bf{W}}^0[n])^{\text H}{\bf N}^0[n].
\end{align}
Furthermore, to represent the channel vector in a standard CS form, let us vectorize the observations ${\mathbf{Y}}^0[n]$, yielding \citep{zhang2022adaptive}
\begin{align}\label{RIScsmodel.eq}
    {\text{vec}}({{\mathbf{Y}}^0}[n]) =& {\text{vec}}(\sqrt{P}({\bf{W}}^0[n])^{\text H}{\mathbf{U}}_{\text{M}}{{\mathbf{H}}_{\text{v}}^0}[n]{\mathbf{U}}_{\text{B}}^{\text H}{{\mathbf{F}}^0[n]}\nonumber\\&+({\bf{W}}^0[n])^{\text H}{\bf N}^0[n])\nonumber\\
    =&[({{\mathbf{F}}^0[n]})^{\text{T}} \otimes ({\bf{W}}^0[n])^{\text H}][({\mathbf{U}}_{\text{B}}^0)^*\otimes{\mathbf{U}}_{\text{M}}^0]{\text{vec}}({{\mathbf{H}}_{\text{v}}^0}[n])\nonumber\\&+{\text{vec}}(({\bf{W}}^0[n])^{\text H}{\bf N}^0[n])\nonumber\\
    =&{\bf \Phi}^0[n]{\bf \Sigma}^0{\bf h}_{{\text v}}^0[n]+{\bf n}^0[n],
\end{align}
where ${\bf \Phi}^0[n]=({{\mathbf{F}}^0[n]})^{\text{T}} \otimes ({\bf{W}}^0[n])^{\text H}$ is the sensing matrix, and ${\bf \Sigma}^0=({\mathbf{U}}_{\text{B}}^0)^*\otimes{\mathbf{U}}_{\text{M}}^0$ is {the overcomplete dictionary or beamspace transformation matrix \citep{9493736,zhang2022adaptive}}, ${\bf h}_{{\text v}}^0[n]$ is the vectorized beamspace channel vector to be estimated, while ${\bf n}^0[n]$ is the noise vector.

Based on (\ref{RIScsmodel.eq}), for the $l$-th, $l=0,...,L-1$, path, where $L$ is the number of physical propagation paths between BS and UE, the residual vector is initialized as $ {{\bf{r}}^0}\left[ n \right] = {\text{vec}}({{\mathbf{Y}}^0}[n])$. Let ${{\mathbf{s}}_i}[n]$ be the $i$-th column of ${\bf S}[n]={\bf \Phi}^0[n]{\bf \Sigma}^0$. Then, the vector atom contributions and AoA/AoD at the BS and the UE can be respectively estimated as 
    \begin{align}\label{atomcontributionRIS1.eq}
        {\widehat i_l} =& \mathop {\arg \max }\limits_{i = 1,...,{N_{\text{M}}}{N_{\text{B}}}} \sum\limits_{n = 0}^{N - 1} {\frac{{\left| {{\mathbf{s}}_i^{\text{H}}[n]{{\mathbf{r}}_{l-1}^0}[n]} \right|}}{{{{\left\| {{{\mathbf{s}}_i}[n]} \right\|}_2}}}},\\
        i_{\text{BR},l}=&\left\lceil {{{\widehat i}_l}/{N_{\text{M}}}} \right\rceil,\\ i_{\text{RM},l}=&\bmod \left({\widehat i_l} - 1,{N_{\text{M}}}\right) + 1,\\
        {\widehat \theta _{\text{BR},l}} =& \arcsin \left(\frac{{{\lambda _{\text{c}}}}}{d}\frac{{{{ i}_{{\text{BR}},l}} - ({N_{\text{B}}} - 1)/2 - 1}}{{{N_{\text{B}}}}}\right),
    \end{align}
 \begin{align}
     {\widehat \phi _{{\text{RM}},l}} =& \arcsin \left(\frac{{{\lambda _{\text{c}}}}}{d}\frac{{{ i}_{{\text{RM}},l} - ({N_{\text{M}}} - 1)/2 - 1}}{{{N_{\text{M}}}}}\right).
 \end{align}
 Afterwards, the basis vector is updated via the orthogonalization on the processing as ${{\bm{\rho }}_l}[n] = {{\mathbf{s}}_{{{\widehat i}_l}}}[n] - \sum\limits_{\widetilde i = 0}^{l - 1} {\frac{{{\mathbf{s}}_{{\widehat i}_ l}^{\text{H}}[n]{{\bm{\rho }}_{\widehat i}}[n]}}{{{{\left\| {{{\bm{\rho }}_{\widehat i}}[n]} \right\|}_2}}}} {{\bm{\rho }}_{\widehat i}}[n]$, while the residual vector for the next iteration is updated to ${{\mathbf{r}}_{l}}[n] = {{\mathbf{r}}_{l - 1}}[n] - {\bm{\alpha} _n}(l){\bm{\rho} _{l}}[n]$,  where ${{\alpha} _n}(l) = \frac{{\bm{\rho} _l^{\text{H}}[n]{{\mathbf{r}}_{l - 1}}[n]}}{{\left\| {{\bm{\rho} _l}[n]} \right\|_2^2}}$.
 To summarize, the DCS-SOMP assisted initial BS precoder and UE combiner design are described in Algorithm \ref{alg:somp_initial}. Note that, \eqref{RIScsmodel.eq} is a standard CS model, many CS algorithms, such as, BP, MP, SBL algorithms, can be employed for the solutions. As mentioned above, in this paper, for the sake of simplicity, the DCS-SOMP algorithm is adopted for its low complexity.
\begin{algorithm}[htbp]
  \caption{DCS-SOMP Assisted Initial Design of BS precoder And UE Combiner}
  \label{alg:somp_initial}
  \begin{algorithmic}[1]
    \Require
     Observation ${\text{vec}}({{\mathbf{Y}}^0}[n])$ ;
     Sensing matrix ${\bf \Phi}^0[n]$;
     Dictionary matrix ${\bf \Phi}^0[n]{\bf \Sigma}^0$.
    \Ensure
      To estimate ${\bm{\widehat \theta} _{\text{BR}}}$, ${\bm{\widehat \phi} _{\text{RM}}}$.
    
    \For{$n=0,...,N-1$}
    \For{$l=0,...,L-1$}
    \State  
    
        ${\widehat i}_l = \mathop {\arg \max }\limits_{i = 1,...,{N_{\text{M}}}{N_{\text{B}}}} \sum\limits_{n = 0}^{N - 1} {\frac{{\left| {{\mathbf{s}}_i^{\text{H}}[n]{{\mathbf{r}}_{l-1}^0}[n]} \right|}}{{{{\left\| {{{\mathbf{s}}_i}[n]} \right\|}_2}}}}$,
        
        $i_{\text{BR},l}=\left\lceil {{{\widehat i}_l}/{N_{\text{M}}}} \right\rceil, i_{\text{RM}}=\bmod \left({\widehat i}_l - 1,{N_{\text{M}}}\right) + 1$,
   \State
   
        ${\widehat \theta _{{\text{BR}},l}} = \arcsin \left(\frac{{{\lambda _{\text{c}}}}}{d}\frac{{{{ i}_{{\text{BR}},l}} - ({N_{\text{B}}} - 1)/2 - 1}}{{{N_{\text{B}}}}}\right)$,
        
        ${\widehat \phi _{{\text{RM}},l}} = \arcsin \left(\frac{{{\lambda _{\text{c}}}}}{d}\frac{{{{ i}_{{\text{RM}},l}} - ({N_{\text{M}}} - 1)/2 - 1}}{{{N_{\text{M}}}}}\right)$.
        \State  Update:

${{\bm{\rho }}_l}[n] = {{\mathbf{s}}_{{{\widehat i}_l}}}[n] - \sum\limits_{\widetilde i = 0}^{l - 1} {\frac{{{\mathbf{s}}_{{\widehat i}_ l}^{\text{H}}[n]{{\bm{\rho }}_{\widehat i}}[n]}}{{{{\left\| {{{\bm{\rho }}_{\widehat i}}[n]} \right\|}_2}}}} {{\bm{\rho }}_{\widehat i}}[n]$

${{\mathbf{r}}_{l}}[n] = {{\mathbf{r}}_{l - 1}}[n] - {\bm{\alpha} _n}(l){\bm{\rho} _{l}}[n]$, where 

${{\alpha} _n}(l) = \frac{{\bm{\rho} _l^{\text{H}}[n]{{\mathbf{r}}_{l - 1}}[n]}}{{\left\| {{\bm{\rho} _l}[n]} \right\|_2^2}}$.
    \EndFor
    \EndFor\\
    \Return ${\bm{\widehat \theta} _{\text{BR},l}}$, ${\bm{\widehat \phi} _{\text{RM},l}}$.
  \end{algorithmic}
\end{algorithm}

After the estimation of the AoDs at BS and the AoAs at UE, {the precoding matrix at BS and the combining matrix at UE remain fixed for the following $J$ blocks, i.e., we set ${\bf F}^1[n]={\bf F}^2[n]=...={\bf F}^J[n]={\bf F}^0[n]$ and ${\bf W}^1[n]={\bf W}^2[n]=...={\bf W}^J[n]={\bf W}^0[n]$. With the aid of these initially designed precoder and combiner, the phase shift matrix $\bm{\Omega}_{t>0}$} can then be designed as detailed in the following subsection. 

\subsection{Stage 2: Parameter Estimation and Phase Shifter Design for RIS}\label{Stage 2: Parameter Estimation and Phase Shifter Design for RIS.subsec}
After the precoding matrices ${\bf F}^1[n]={\bf F}^2[n]=...={\bf F}^J[n]$ at BS and the combining matrix ${\bf W}^1[n]={\bf W}^2[n]=...={\bf W}^J[n]$ at UE for the blocks $t>0$ are obtained, as shown in the last subsection, the AoR at RIS can then be estimated using the time block of $t>0$, as shown in Fig.~\ref{fig:CEdesign}, which can be further used in the localization of UE.

More specifically, as mentioned in the last subsection, after the estimation of $\theta_{{\text {BR}},l}$ and $\phi_{{\text{RM}},l}$ in the first stage, the corresponding beam training matrices at BS and UE for RIS can be designed as follows:
\begin{align}\label{beamdesign.eq}
    {\bf{W}}^t[n]\approx& \mathbf{A}_{\text{M}}(\widehat{\bm\phi}_{\text {RM}}),\nonumber\\
    {\bf{F}}^t[n]\approx& \mathbf{A}_{\text{B}}(\widehat{\bm\theta}_{\text {BR}}).
\end{align}
where $t=1,...,J$, and $n=0,...,N-1$. 

For the time block $t=0$, we note that the number $G$ of symbol durations, as included by the red columns in Fig.~\ref{fig:CEdesign} is much larger than the number of paths $L$, i.e. $G\gg L$. By contrast, for $t>0$, as shown by the yellow columns in Fig.~\ref{fig:CEdesign}, for each time block, the number of symbol durations is only required to be the number of paths, making ${\bf{W}}^t[n]\in \mathbb{C}^{N_{\text B}\times{L_{\text{RM}}} }$ and ${\bf{F}}^t[n]\in \mathbb{C}^{N_{\text B}\times{L_{\text{BR}}} }$, {where $L_{\text{RM}}$ denotes the number of paths between RIS and UE, and $L_{\text{BR}}$ of that between BS and RIS, respectively.} In this regard, the training overhead can be significantly reduced for the $t>0$ blocks. 

Following (\ref{simplifiedreceivermodel.eq}) and (\ref{beamdesign.eq}), the received signal at UE corresponding to $t>0$ time blocks can be derived as
\begin{align}\label{simplifiedreceivedsignal.eq}
     {\bf Y}^t[n]=& ({\bf{W}}^t[n])^{\rm H}{\mathbf{H}}^t[n]{{\mathbf{F}}^t[n]}+({\bf{W}}^t[n])^{\text H}{\bf N}^t[n]\nonumber\\
    =& ({\bf{W}}^t[n])^{\rm H}\mathbf{A}_{\text{M}}(\bm{\phi}_{\text {RM}}){\mathbf{H}}_{\text{eff}}^t[n]\mathbf{A}_{\text{B}}(\bm{\theta}_{\text {BR}}){{\mathbf{F}}^t[n]}\nonumber\\&+({\bf{W}}^t[n])^{\text H}{\bf N}^t[n]\nonumber\\
    \approx&{\bf H}_{\text{eff}}^t[n]+({\bf{W}}^t[n])^{\text H}{\bf N}^t[n], t=1,...J;n=0,...,N-1,
\end{align}
where we applied ${\bf{W}}^t[n]^{\rm H}\mathbf{A}_{\text{M}}({\bm \phi}_{\text {RM}})\approx \bf{I}$ and $\mathbf{A}_{\text{B}}({\bm \theta}_{\text {BR}}){{\mathbf{F}}^t[n]}\approx\bf{I}$, owing to the precoder and combiner designed for BS and UE as shown in \eqref{beamdesign.eq}, and the effective channel ${\mathbf{H}}_{\text{eff}}^t[n]$ is defined in (\ref{effectivechannel.eq}). It can be shown that the $(a,b)$-th entry of the effective channel ${\bf H}_{\text{eff}}^t[n]$ can be expressed as
\begin{align}\label{recallingchannel.eq}
   [{\bf H}_{\text{eff}}^t[n]]_{(a,b)} = ({\widehat{\bm \rho}}_{\rm{RM}})_a{\bm\omega}_t^{\rm T}{\bf a}([{\bm \theta}_{\rm{diff}}]_{(a,b)})({\widehat{\bm \rho}}_{\rm{BR}})_b,
\end{align}
where ${\bm\omega}_t$ is a vector denoting the diagonal elements of the phase shift matrix ${\bm\Omega}_t$, and ${\bm \theta}_{\rm{diff}}={\rm{asin}}({\rm{sin}}([{\bm\phi}_{{\rm{BR}}}]_b)-{\rm{sin}}([{\bm\theta}_{\rm{RM}}]_a))$ is the spatial frequency to avoid the angle ambiguity when recovering the angles \citep{he2021channel}. Now let us vectorize \eqref{effectivechannel.eq}, the effective channel vector can be represented as ${\bf h}_{\rm{eff}}^t[n] = {\rm{vec}}({\bf H}_{\text{eff}}^t[n]) \in \mathbb{C}^{L_{\rm{RM}} L_{\rm{BR}}\times1}$, and the elements of ${\bf h}_{\rm{eff}}^t[n]$ can be written as 
\begin{align}\label{veceffchannel.eq}
    [{\bf h}_{\rm{eff}}^t[n]]_k = {\widehat{\bm{\rho}}}_k[n]{\bm\omega}_t{\bf a}({\widehat{\bm \theta}}_{{\rm{diff}},k}), k=1,...,L_{\rm{RM}} L_{\rm{BR}},
\end{align}
where ${\widehat{\bm{\rho}}}_k[n]=({\widehat{\bm \rho}}_{\rm{RM}}[n])_a({\widehat{\bm \rho}}_{\rm{BR}}[n])_b$, ${\widehat{\bm \theta}}_{{\rm{diff}},k}=\arcsin(\sin([{\bm\phi}_{\rm{BR}}]_b)-\sin([{\bm\theta}_{{\rm{RM}}}]_a))$, while $a=\mod({k-1,L_{\rm{RM}}})+1$ and $b=\lceil k/L_{\rm{RM}}\rceil$.
Consequently, when considering all the observations over $J$ blocks, according to \eqref{simplifiedreceivedsignal.eq}, we have
\begin{align}\label{gatheredY.eq}
    {\bf{Y}}[n]=&[{\rm{vec}}({\bf Y}^1)[n],{\rm {vec}}({\bf Y}^2)[n],...,{\rm {vec}}({\bf Y}^J)[n]],\nonumber\\&n=0,...,N-1.
\end{align}
Correspondingly, the effective channel is ${\widehat{\bf{H}}}_{\text{eff}}[n]=[{\bf{h}}_{\text{eff}}^1[n],{\bf{h}}_{\text{eff}}^2[n],...,{\bf{h}}_{{\text{eff},}k}^J[n]]$. Furthermore, based on (\ref{simplifiedreceivedsignal.eq}), (\ref{recallingchannel.eq}) and \eqref{veceffchannel.eq}, it can be shown that the $k$-th row of ${\bf{Y}}[n]$ is 
\begin{align}\label{csreceivedsignal.eq}
    {\bf{Y}}_{k,:}[n]\approx&{\widehat{\bf{H}}}_{\text{eff},{k,:}}[n]^{\rm {T}}+{\widehat{\bf{N}}}_{k,:}[n]\nonumber\\
    =&{\widehat{\bm{\Omega}}}{\widehat{\bm{\rho}}}_k[n]\mathbf{a}({\widehat{\bm \theta}}_{{\rm{diff}},k})+{\widehat{\bf{N}}}_{k,:}[n], k=1,...,L_{\rm{RM}} L_{\rm{BR}}.
\end{align}
In (\ref{csreceivedsignal.eq}), ${\widehat{\bm{\Omega}}}=[{\bm\omega}^1,{\bm\omega}^2,...,{\bm\omega}^J]^{\rm{T}}$ is from the phase shift vectors used for the transmissions over $J$ blocks, which are random phases, as mentioned in Section \ref{problemformulationpaper3.sec}. ${\widehat{\bf{N}}}[n]=[{\rm{vec}}({\bf{W}}^1[n]^{\text H}{\bf N}^1[n]),{\rm{vec}}({\bf{W}}^2[n]^{\text H}{\bf N}^2[n]),...,{\rm{vec}}({\bf{W}}^J[n]^{\text H}{\bf N}^J[n])]$ is the stacked AWGN matrix, where the covariance matrix of the AWGN vector with respect to the $t$-th block is ${\bf R}_t = \sigma^2{\bf{W}}^t[n]^{\text H}{\bf{W}}^t[n]$. Moreover, the covariance matrix ${\bf R}\in \mathbb{C}^{TL_{\rm{RM}}L_{\rm{BR}}\times TL_{\rm{RM}}L_{\rm{BR}} }$ of the vectorized noise matrix ${\text{vec}}({\widehat{\bf{N}}}[n])$ is a block diagonal matrix, with the matrices ${\bf R}_t$ on its diagonal.

Our objective is to estimate the AoR ${\bm\theta}_{\rm{RM}}$ and the ToA ${\bm\tau}_{\rm{RM}}$ between RIS and UE. Thus, the problem can be formulated as the estimation problem for ${\widehat{\bm{\rho}}}_k[n]$ and ${\widehat{\bm \theta}}_{{\rm{diff}},k}$, since the locations of both BS and RIS are usually fixed in most applications in practice, and hence the related angles such as ${\bm\theta}_{\rm{BR}}$, ${\bm\phi}_{\rm{BR}}$, and the ToA ${\bm\tau}_{\rm{BR}}$ can be pre-measured \citep{he2020adaptive}. Let us introduce the DFT matrix $\bf{U}_{\rm{R}}$ for the sparse formulation of $\mathbf{a}({\widehat{\bm \theta}}_{{\rm{diff}},k})$ in (\ref{csreceivedsignal.eq}). Hence $\bf{U}_{\rm{R}}$ is defined in the same way as (\ref{beamspaceRIS1.eq}). Then, the received signal of (\ref{csreceivedsignal.eq}) can be alternatively expressed as
\begin{align}\label{csreceivedsignal2.eq}
    {\bf{Y}}_{k,:}[n]=\bm{\Psi}{\bf{h}}_k[n]+{\widehat{\bf{n}}}[n],
\end{align}
where $\bm{\Psi}={\widehat{\bm{\Omega}}}_k^{\rm{T}}\bf{U}_{\rm{R}}$ can be explained as a sensing matrix, and ${\bf{h}}_k[n]$ is the desired sparse vector, which embeds the location information AoR and ToA. 

It can be understood from \eqref{csreceivedsignal2.eq} that the received signals have three dimensions, namely time blocks, subcarriers, and the possible propagation paths, which in total have the number of $L_{\rm{RM}} L_{\rm{BR}}$. However, for localization purpose, only the path with the highest power is effective, which can be extracted from \eqref{csreceivedsignal.eq} via the power measurement, expressed as 
\begin{align}\label{maxpath.eq}
\widehat k = \arg \max {\left\| {{{\bf{Y}}_{k,:}}\left[ n \right]} \right\|_2^2}.
\end{align}
Consequently, when considering the $J$ blocks, $N$ subcarriers, and $\widehat k$-th path components, corresponding to \eqref{csreceivedsignal2.eq}, the received signals for localization can be expressed as
\begin{align}\label{stackedY.eq}
    {\widehat {\bf Y}} = {\bm{\Psi}}{\widehat{\bf H}} + {\bf Z},
\end{align}
where ${\widehat {\bf Y}}\in \mathbb{C}^{T\times N}$, ${\bm{\Psi}}\in \mathbb{C}^{T\times N_{\rm R}}$ and ${\widehat{\bf H}}\in \mathbb{C}^{N_{\rm R}\times N}$, while each column of ${\bf Z}$ is the AWGN. Finally, after the vectorization of \eqref{stackedY.eq}, the received signal can be represented {in} a group sparse format as
\begin{align}\label{vecstackedY.eq}
    {\widehat {\bf y}} = {\widehat{\bm{\Psi}}}{\widehat{\bf h}} + {\bf z}.
\end{align}
In \eqref{vecstackedY.eq}, ${\widehat {\bf y}}={\rm{vec}({\widehat {\bf Y}}^{\rm T})}\in \mathbb{C}^{NJ \times 1}$,${\widehat{\bm{\Psi}}}=(\bm{\Psi} \otimes {\bf I}_N)\in \mathbb{C}^{NJ\times NN_{\rm R}}$, and ${\widehat{\bf h}}={\rm{vec}({\widehat {\bf H}}^{\rm T})}\in \mathbb{C}^{NN_{\rm R} \times 1}$. Furthermore, according to \eqref{csreceivedsignal.eq}, the noise variance for the $t$-th time block and the $n$-th subcarrier can be expressed as $\widehat{\sigma}_t= \sigma^2{\bf{W}}_{\widetilde{k},:}^t[n]^{\text H}{\bf{W}}_{\widetilde{k},:}^t[n]$. Hence, the covariance of the elements in $\bf z$ can be written as $\widehat{\bf R}_{t,t}=\widehat{\sigma}_t$, {where $\widehat{\bf R}_{t,t}$ denotes the $t$-th diagonal element of $\widehat{\bf R}$}. Therefore, in \eqref{vecstackedY.eq}, the covariance matrix of the noise vector $\bf z$ can be expressed as $\widetilde {\bf R} = \mathbb{E}\left\{ {\bf z}{\bf z}^{\rm H} \right\}=(\widehat{\bf R}\otimes {\bf I}_N)$.

Based on \eqref{vecstackedY.eq}, the group sparse channel vector can be estimated. Then, the AoR and ToA can be extracted from the estimated channel vector, from which the location of UE can be estimated, as detailed in the next section.

\section{Proposed Group Sparse Bayesian Learning For UE Localization}\label{proposedSBL.sec}
\subsection{Conventional Group SBL}\label{coventionalGSBL.subsec}
The desired channel vector ${\widehat{\bf h}}$, which is sparse in both time and spatial domain, and contains the information of AoR and ToA for localization, is group-sparse. Therefore, we have the $\textit{a~prior}$ about the desired channel vector ${\widehat{\bf h}}$ as 
\begin{align}\label{aprior.eq}
p\left( {\widehat {\bf{h}};{\bm\Gamma} ,{\bf{M}}} \right) = \prod\limits_{j = 1}^{{N_R}} {p\left( {\bf{h}}_j;{\gamma _j},{\bf{M}} \right)},
\end{align}
where $\bf M$ denotes the correlation matrix of ${\bf{h}}_j$, which is assumed to be unknown along with the hyperparameter ${\gamma _j}$, ${\bf{h}}_j={\widehat{\bf h}}[(j-1)N+1:jN]$ denotes the $j$-th group of ${\widehat{\bf h}}$, which for given $\gamma _j$ and ${\bf{M}}$ obeys the distribution of
\begin{align}\label{aprior2.eq}
p\left( {{{\bf{h}}_j};{\gamma _j},{\bf{M}}} \right) = \frac{1}{{{{\left( {\pi {\gamma _j}} \right)}^N}\det \left( {\bf{M}} \right)}}\exp \left( { - \frac{{{\bf{h}}_j^{\rm{H}}{{\bf{M}}^{ - 1}}{{\bf{h}}_j}}}{{{\gamma _j}}}} \right),
\end{align}
According to the automatic relevance determination \citep{tipping2001sparse}, ${\bf{h}}_j$ can be generated based on the hyperparameter ${\gamma}_j$. When considering all the $N_{\rm R}$ groups, we have a vector $\bm{\gamma}$ controlling the prior variance of the group of elements in ${\bf{h}}$. Let ${\bm\Gamma}={\rm{diag}}(\bm{\gamma}) \in \mathbb{R}^{N_{\rm R}\times N_{\rm R}}$ be a diagonal matrix with the hyperparameter vector $\bm{\gamma}$ on its diagonal. According to \citep{zhang2011sparse}, the SBL-based algorithms consist of two stages, namely the expectation and maximization, which are termed as E-step and M-step. Specifically, to solve our problem, the expected value for the $(iter-1)$-th iteration of the log-likelihood function ${\ell}({\bm{\Gamma}},{\bf{M}}|{\widehat{\bm\Gamma}_{iter-1}},\widehat{\bf{M}}_{iter-1})$, corresponding to the complete data $\{{\widehat {\bf y}},{\widehat{\bf h}}\}$, is determined by the E-step as 
\begin{align}\label{E-step.eq}
    {\ell}({\bm{\Gamma}},&{\bf{M}}|{\widehat{\bm\Gamma}_{iter-1}},\widehat{\bf{M}}_{iter-1}) \nonumber\\
    &= \mathbb{E}_{{\widehat{\bf h}}|{\widehat {\bf y}};{\widehat{\bm\Gamma}_{iter-1}},\widehat{\bf{M}}_{iter-1}}\{{\text{log}p({\widehat {\bf y}},{\widehat{\bf h}};{\bm{\Gamma}},{\bf{M}})}\}.
\end{align}

As ${\text{log}}p({\widehat {\bf y}}|{\widehat{\bf h}})$ is independent of the hyperparameter matrix $\bm{\Gamma}$ and the correlation matrix $\bf{M}$, By employing Bayes' rule on \ref{E-step.eq}, the M-step for maximizing the log-likelihood function ${\ell}({\bm{\Gamma}},{\bf{M}}|{\widehat{\bm\Gamma}_{iter-1}},\widehat{\bf{M}}_{iter-1})$ to update ${\bm{\Gamma}}$ and ${\bf{M}}$ can be expressed as
\begin{align}\label{M-step1.eq}
    ({\widehat{\bm\Gamma}_{iter}},\widehat{\bf{M}}_{iter})=\mathop {\arg \max }\limits_{{\bm{\Gamma}},{\bf{M}}} \mathbb{E}\{{\text{log}p({\widehat{\bf h}};{\bm{\Gamma}},{\bf{M}}})\}.
\end{align}
Then, based on \eqref{aprior.eq} and \eqref{aprior2.eq}, each $\gamma_j$ is decoupled by the M-step, and according to \citep{zhang2011sparse}, the \textit{a~posteriori} probability density function (PDF) for the $iter$-th iteration is given by $p\left( {\widehat {\bf{h}}|{\bf{y}};{{\widehat {\bm\Gamma} }_{iter - 1}},{\widehat{\bf{M}}_{iter - 1}}} \right)\sim{\mathcal{CN}}\left( \bm{\mu }_{iter},{\bm\Sigma} _{iter} \right)$ with
\begin{align}\label{aposteriorievaluation.eq}
    \bm{\mu }_{iter} =& {\bm\Sigma} _{iter}{\widehat{\bm\Psi}}^{\rm{H}}{\widetilde{\bf R}}^{-1}{\widehat{\bf y}}, \nonumber\\
    {\bm\Sigma} _{iter} =& \left[({\widehat{\bm\Gamma}} _{iter-1}\otimes {\widehat{\bf{M}}_{iter - 1}})^{-1}+{\widehat{\bm\Psi}}^{\rm{H}}{\widetilde{\bf R}}^{-1}{\widehat{\bm\Psi}}\right]^{-1}.
\end{align}
Then, the estimate to the hyperparamter $\gamma _{j,iter}$ can be updated as \citep{srivastava2021sparse}
\begin{align}\label{updatedhyper.eq}
    \gamma _{j,iter}=\frac{1}{N}{\rm{Trace}}((\widehat{\bf{M}}_{iter-1})^{-1}({\widehat{\bm\Gamma}} _{iter,j}+\bm{\mu }_{iter,j}(\bm{\mu }_{iter,j})^{\rm H})),
\end{align}
where ${\bm\mu} _{j,iter}={\bm\mu} _{j}[(j-1)N+1:jN]\in \mathbb{C}^{N\times 1}$ and ${\bm\Sigma} _{iter,j}={\bm\Sigma} _{iter}[(j-1)N+1:jN,(j-1)N+1:jN]\in\mathbb{C}^{N\times N}$ denote respectively the \textit{a~posteriori} mean and covariance of the channel vector ${\bf{h}}_j$ of the $j$-th group in ${\widehat{\bf h}}$. Simultaneously, the correlation matrix is updated as \citep{srivastava2021sparse}
\begin{align}\label{updatecorrelation.eq}
    {\widehat {\bf{M}}_{iter}} = \frac{1}{{{N_{\rm R}}}}\sum\limits_{iter = 1}^{{N_{\rm R}}} {\frac{{\widehat{\bm\Gamma}} _{iter,j}+\bm{\mu }_{iter,j}(\bm{\mu }_{iter,j})^{\rm H}}{{{\gamma _{j,iter}}}}}.
\end{align}
At the end, when the estimation converges, the estimation of the desired channel vector ${\widetilde{\widehat{\bf h}}}$ is given by the converged \textit{a~posteriori} mean ${\bm\mu }_{iter}$. 

To estimate the ToA from the multi-carrier signals, the group-sparse channel vector ${\widetilde{\widehat{\bf h}}}$ is estimated on the subcarrier-by-subcarrier basis, expressed as ${\widetilde{\widehat{\bf H}}}={\rm{reshape}}({\widetilde{\widehat{\bf h}}},N)\in \mathbb{C}^{N_{\rm{R}} \times N}$. Then, the estimated channel gain ${\widetilde{\widehat{\bm{\rho}}}}_{\widehat k}$ and spatial frequency ${\widetilde{\widehat{\bm \theta}}}_{{\rm{diff}},{\widehat k}}$ for the $\widehat k$-th path can be expressed as
\begin{align}\label{angleestimation.eq}
    {\widetilde{\widehat{\bm{\rho}}}}_{\widehat k}=&{\rm{max}({\widetilde{\widehat{\bf H}}},2)}\in \mathbb{C}^{1 \times N},\nonumber\\
    {\widehat p} =& \mathop {\arg \max }\limits_{p = 1,...,{N_R}} \sum\limits_{v = 1}^N {{{\widetilde{\widehat{\bf H}}}_{:,v}}},\nonumber\\
    {\widehat\theta _{{\text{diff}},{\widehat k}}} =& \arcsin \left(\frac{{{\lambda _{\text{c}}}}}{d}\frac{{{\widehat p} - ({N_{\text{R}}} - 1)/2 - 1}}{N_{\text{R}}}\right).
\end{align}

Based on \eqref{veceffchannel.eq} and \eqref{angleestimation.eq}, as the locations of BS and RIS are assumed to be fixed, the AoA at RIS $\phi_{{\rm{BR}},{\widehat k}}$ can be assumed to be known, which can be pre-measured when RIS is deployed. Thus, the AoR at RIS can be estimated as \citep{he2021channel}
\begin{align}\label{GSBLAoR.eq}
    {\widehat\theta _{{\text{RM}},{\widehat k}}}= \arcsin(\sin(\phi_{{\rm{BR}},{\widehat k}})- \sin({\widehat\theta _{{\text{diff}},{\widehat k}}})).
\end{align}

The ToA is related to the channel gain in OFDM systems. Let us define ${\mathbf{g}}({\tau _{{\rm{BRM}},{\widehat k}}})={\left[ {1,...,{e^{ - j2\pi (N - 1){\tau _{{\rm{BRM}},{\widehat k}}}/(N{T_s})}}} \right]^{\text{T}}}$ as one of the grid candidates found from the time between BS and UE. Then, the ToA between BS and UE can be estimated as
\begin{align}\label{timedelayRIS.eq}
{\widehat \tau _{{\rm{BRM}},{\widehat k}}} = \mathop {\arg \max }\limits_{{\tau _{{\rm{BRM}},k}}} {\left| {{\mathbf{g}}({\tau _{{\rm{BRM}},{\widehat k}}}){\widetilde{\widehat{\bm{\rho}}}}_{\widehat k}} \right|^2},
\end{align}
where ${\widetilde{\widehat{\bm{\rho}}}}_{\widehat k} = {\left[ {{\widetilde{\widehat{{\rho}}}}_{\widehat k}\left[ 0 \right],...,{\widetilde{\widehat{{\rho}}}}_{\widehat k}\left[ {N - 1} \right]} \right]^\text T}$ holds the finally estimated channels of the $N$ subcarriers.
Ultimately, when considering that the above delay consists of the delay between BS and RIS and that between RIS and UE, the ToA $\tau_{{\rm{RM}},{\widehat k}}$ between RIS and UE can be obtained as
\begin{align}\label{AoDToA.eq}
    \widehat \tau_{{\rm{RM}},{\widehat k}}=&{\widehat \tau _{{\rm{BRM}},{\widehat k}}}-\tau_{{\rm{BR}},{\widehat k}}.
\end{align}
Finally, with the aid of the estimated ${\widehat\theta _{{\text{RM}},{\widehat k}}}$ in \eqref{angleestimation.eq} and ${\widehat \tau _{{\rm{BRM}},{\widehat k}}}$ of \eqref{timedelayRIS.eq}, the location of the UE can be estimated as 
\begin{align}\label{MSpositionrecovery.eq}
    \widehat{\bf{m}}={\bf{r}}+c\widehat \tau_{{\rm{RM}},{\widehat k}}[\cos({\widehat \theta_{\text {RM},{\widehat k}}}),\sin({\widehat \theta_{\text {RM},{\widehat k}}})]^{\rm{T}}.
\end{align}

To sum up, the RIS-aided localization scheme based on the GSBL algorithm, which is operated during the second stage, as discussed in Section~\ref{Stage 2: Parameter Estimation and Phase Shifter Design for RIS.subsec}, can be summarized as Algorithm \ref{alg:GSBL_stage2}.

\begin{algorithm}[htbp]
  \caption{GSBL algorithm for the RIS-Aided Localization in mmWave System}
  \label{alg:GSBL_stage2}
  \begin{algorithmic}[1]
    \Require
     Observations ${\widehat{\bf{y}}}$ ;
     Sensing matrix $\widehat{\bm{\Psi}}$;
     Maximum iteration $K_{\rm{max}}$.
    \Ensure
      To estimate ${\widehat \theta_{\text {RM},{\widehat k}}}$, $\widehat \tau_{{\rm{RM}},{\widehat k}}$.

    \State Initialize hyperparameters: ${\widehat{\bm\Gamma}} _{0}={\bf I}_{N_{\rm R}}$, ${\widehat{\bm\Gamma}} _{-1}={\bf 0}_{N_{\rm R}}$
    \For{$iter=0,...,K_{\rm{max}}$}
    \State  Expectation: Evaluate the \textit{a~postriori} mean ${\bm \mu}_{iter}$ and covariance matrix ${\bm\Sigma} _{iter}$ according to \eqref{aposteriorievaluation.eq}.
    
   \State Maximization: Update the hyperparameters $\gamma _{j,iter}$ based on \eqref{updatedhyper.eq}, and the correlation matrix ${\widehat {\bf{M}}_{iter}}$ based on \eqref{updatecorrelation.eq}, with ${\widehat{\bm\Gamma}} _{iter}={\rm{diag}}(\gamma _{1,iter},\gamma _{2,iter},...,\gamma _{N_{\rm R},iter})$.
   \EndFor 
   \State  AoR and ToA recovery based on \eqref{angleestimation.eq} to \eqref{AoDToA.eq}. \\
   \Return ${\widehat \theta_{\text {RM},k}}$, $\widehat \tau_{{\rm{RM}},k}$.
  \end{algorithmic}
\end{algorithm}

\subsection{Proposed Localization Scheme Using Modified Temporally Correlated Multiple Sparse Bayesian Learning}\label{Proposed Localization Scheme Using Modified TMSBL.subsec}
The GSBL algorithm jointly considers $N$ subcarriers, $J$ time blocks and $N_{\rm R}$ grids/antennas of RIS, leading to the high complexity of the inversion of ${\widetilde{\bf R}}\in \mathbb{C}^{JN\times JN}$ and ${\bm{\Sigma}} _{iter}\in \mathbb{C}^{N_{\rm{R}}N \times N_{\rm{R}}N}$ in \eqref{aposteriorievaluation.eq}. In order to reduce the size of these matrices in \eqref{aposteriorievaluation.eq} and hence the implementation complexity, inspired by (26) in \citep{zhang2011sparse}, the covariance matrix ${\bm\Sigma} _{iter}$ in \eqref{aposteriorievaluation.eq} can be approximated as 
\begin{align}\label{sigmaapprox.eq}
    {\bm\Sigma} _{iter} =& [({\widehat{\bm\Gamma}} _{iter-1}\otimes {\widehat{\bf{M}}_{iter - 1}})^{-1}+{\widehat{\bm\Psi}}^{\rm{H}}{\widetilde{\bf R}}^{-1}{\widehat{\bm\Psi}}]^{-1}\nonumber\\
    =&[({\widehat{\bm\Gamma}} _{iter-1}\otimes {\widehat{\bf{M}}_{iter - 1}})^{-1}+({{\bm\Psi}}^{\rm{H}}{\widehat{\bf R}}^{-1}{{\bm\Psi}})\otimes {\bf I}_N]^{-1}\nonumber\\
    \approx&(({\widehat{\bm\Gamma}} _{iter-1})^{-1}+{{\bm\Psi}}^{\rm{H}}{\widehat{\bf R}}^{-1}{{\bm\Psi}})^{-1}\otimes {\widehat{\bf{M}}_{iter - 1}}.
\end{align}
From \eqref{sigmaapprox.eq}, ${\bm\Sigma} _{iter}$ can be constructed as ${\bm\Sigma} _{iter,j}=[{\widehat{\bm\Sigma}} _{iter}]_{j,j}{\widehat{\bf{M}}_{iter - 1}}$, where ${\widehat{\bm\Sigma}}=(({\widehat{\bm\Gamma}} _{iter-1})^{-1}+{\widehat{\bm\Psi}}^{\rm{H}}{\widetilde{\bf R}}^{-1}{\widehat{\bm\Psi}})^{-1}$, and ${\bm\Sigma} _{iter,j}$ is the $j$-th diagonal element. With the aid of the approximation of \eqref{sigmaapprox.eq}, according to (23) and (24) in \citep{zhang2011sparse}, the mean $\bm{\mu }_{iter}$ can then be approximated as
\begin{align}\label{approxmu.eq}
    \bm{\mu }_{iter} =& {\bm\Sigma} _{iter}{\widehat{\bm\Psi}}^{\rm{H}}{\widetilde{\bf R}}^{-1}{\widehat{\bf y}}\nonumber\\
    =&{\bm\Sigma} _{iter}[({{\bm\Psi}}^{\rm{H}}{\widehat{\bf R}}^{-1})\otimes{\bf I}_N]{\widehat {\bf y}}\nonumber\\
    \approx&[({\widehat{\bm\Sigma}} _{iter}{{\bm\Psi}}^{\rm{H}}{\widehat{\bf R}}^{-1})\otimes{\bf I}_N]{\rm{vec}({\widehat {\bf Y}}^{\rm T})}\nonumber\\
    =& {\rm{vec}}([{\widetilde{\widehat{\bf H}}}_{iter}]^{\rm T}),
\end{align}
where ${\widetilde{\widehat{\bf H}}}_{iter} = {\widehat{\bm\Sigma}} _{iter}{{\bm\Psi}}^{\rm{H}}{\widehat{\bf R}}^{-1}{\widehat {\bf Y}}$. Then, when substituting the approximations of \eqref{sigmaapprox.eq} and \eqref{approxmu.eq} into \eqref{aposteriorievaluation.eq}, the hyperparameter $\gamma _{j,iter}$ can be updated as
\begin{align}\label{TMSBLgamma.eq}
    \gamma _{j,iter}=[{\widehat{\bm\Sigma}} _{iter}]_{j,j}+\frac{1}{N}([{\widetilde{\widehat{\bf H}}}_{iter}]_{j,:})^{\rm H}(\widehat{\bf{M}}_{iter-1})^{-1}[{\widetilde{\widehat{\bf H}}}_{iter}]_{j,:},
\end{align}
associated with the correlation matrix ${\widehat {\bf{M}}_{iter}}$ represented as
\begin{align}\label{TMSBLcorrelation.eq}
    {\widehat {\bf{M}}_{iter}}=&\frac{1}{N_{\rm R}}\times\nonumber\\
    &\left[\sum\limits_{j = 1}^{N_{\rm R}} {\frac{[{\widehat{\bm\Sigma}} _{iter}]_{j,j}}{\gamma _{j,iter}}}+\sum\limits_{j = 1}^{N_{\rm R}}{\frac{1}{\gamma _{j,iter}}}[{\widetilde{\widehat{\bf H}}}_{iter}]_{j,:}([{\widetilde{\widehat{\bf H}}}_{iter}]_{j,:})^{\rm H}\right].
\end{align}
Note that, in \eqref{TMSBLcorrelation.eq}, the ambiguity between ${\widehat {\bf{M}}_{iter}}$ and $\gamma_j$ is unable to be removed. Furthermore, it is not robust in low SNR region due to the errors introduced from the unreliable estimates of $\gamma_j$ and $[{\widetilde{\widehat{\bf H}}}_{iter}]_{j,:}$. To increase the robustness, \eqref{TMSBLcorrelation.eq} can be updated according to the formulas \citep{zhang2011sparse}
\begin{align}\label{TMSBLcorrelation2.eq}
    {\widetilde {\bf{M}}_{iter}}=&\sum\limits_{j = 1}^{N_{\rm R}}{\frac{1}{\gamma _{j,iter}}}[{\widetilde{\widehat{\bf H}}}_{iter}]_{j,:}([{\widetilde{\widehat{\bf H}}}_{iter}]_{j,:})^{\rm H}+\kappa{\bf I}_N,\nonumber\\
    {\widehat {\bf{M}}_{iter}}=&\frac{{\widetilde {\bf{M}}_{iter}}}{{{{\left\| {\widetilde {\bf{M}}_{iter}} \right\|}_F}}},
\end{align}
where the constant $\kappa=2$ \citep{zhang2011sparse}. 

In summary, the proposed two-stage localization scheme employs the conventional estimation for OFDM mmWave channel in the first stage associated with a randomly generated phase shift matrix at RIS in the beamspace domain, to design the precoder of BS and combiner of UE. Then, during the second stage, multiple time blocks with time-domain sparsity are utilized to estimate the AoR at RIS and the ToA at UE from RIS, from which the location is recovered by UE, according to the down-link channel parameters estimated. 

\section{Performance Results and Analysis}\label{Simulationsetup.sec}
\subsection{Simulation Setup}
\begin{table}[htbp]
\centering
\caption{Simulation parameters}
\begin{tabular}{|c|c|}
\hline
Parameters           & Value                                                   \\ \hline
No. of Antennas $N_{\text{b}}$, $N_{\text{m}}$, $N_{\text{r}}$ & 8 \citep{10285510} or 16 \citep{wang2021joint, he2020adaptive}               \\ \hline
Speed of Light $c$       & $2.99792\times{10^8}$ m/s \\ \hline
Carrier Frequency $f_c$   & 60~GHz                                                   \\ \hline
Bandwidth $B$          & 100~MHz \citep{shahmansoori2017position,he2020adaptive}                                                  \\ \hline
Path Loss of LoS path $\rho_{{\rm{BR}},0}$           & ${\rho _{{\rm{BR,}}0}} = \frac{{\lambda }}{{4\pi {d_{{\rm{BR}},0}}}}$ \citep{wirelessthroughreconfigurable}                                               \\ \hline
Path Loss of NLoS path $\rho_{{\rm{BR}},l}$           & ${\rho _{{\rm{BR,}}l}} = \frac{{\sqrt{\varsigma}\lambda }}{{4\pi {d_{{\rm{BR}},l}}}}$ \citep{wirelessthroughreconfigurable}                                              \\ \hline
Reflection Loss $\varsigma $     & -13~dB \citep{wang2021joint}                                                   \\ \hline
No. of Pilot Subcarriers $N$ & 10                                                     \\ \hline
Location of BS, $\mathbf{b}$      & $[0,0]^\text T$                                        \\ \hline
Location of RIS, $\mathbf{r}$      & $[2.5,4]^\text T$                                               \\ \hline
Location of UE, $\mathbf{m}$      & $[5,3]^\text T$                                             \\ \hline
Locations of the first scatterer $\mathbf{s}_1$                         & $[1,3]^\text T$                  \\ \hline
Locations of the second scatterer $\mathbf{s}_2$                         & $[4,2]^\text T$                  \\ \hline
\end{tabular}
\label{tab:parameters}
\end{table}                 
In this section, we demonstrate the performance of the RIS-aided localization by investigating the effect from the different perspectives, such as different channel estimation algorithms, the number of training blocks, and the position of RIS. As detailed in Table~\ref{tab:parameters}, we consider a system as shown in Fig.~\ref{fig:systemmultipleRISs}, with BS, UE and RIS all employing 8 antennas \citep{10285510} (or 16 antennas when investigating the impact of different number of antennas \citep{wang2021joint, he2020adaptive}). The number of time blocks is set to $J=64$, while the number of pilot subcarriers is $N=10$, unless specified. As shown in Table \ref{tab:parameters}, the locations of BS, UE, and RIS are at ${\bf{b}}=[0,0]^\text T$, ${\bf{m}}=[5,3]^\text T$, and ${\bf{r}}=[2.5,4]^\text T$, respectively. The carrier frequency $f_c$ is set to 60~GHz, while the bandwidth is $B$ = 100~MHz \citep{shahmansoori2017position,he2020adaptive}\footnote{In this paper, we assume a bandwidth of 100~MHz as a design example. A wider bandwidth is also applicable, provided that $B \ll {f_c}$.}. The signal-to-noise ratio (SNR) is defined as $P/{\sigma}^2$. The reflection loss is set to $\varsigma=-13$~dB \citep{wang2021joint}. Finally, the estimation performance is measured by the root mean squared error (RMSE), defined as:
\begin{align}
   \text {RMSE} = \sqrt {\frac{1}{K}{{\sum\limits_{k=1}^K {\left\| {\widehat {\bf{q}} - {\bf{q}}} \right\|}_2^2 }}},
\end{align}
where $K$ denotes the number of Monte Carlo trials, ${\bf{q}}$ and $\widehat {\bf{q}}$ are the true and estimated UE location or orientation. In addition to the above-mentioned, there are also some other parameters specified in Table~\ref{tab:parameters}.

\textbf{Benchmark:} In this paper, we compare the RMSE performance of the proposed modified TMSBL algorithm with the OMP \citep{wei2021channel}, DCS-SOMP \citep{shahmansoori2017position}, AMP \citep{kim2011belief}, SBL \citep{9493736} and the GSBL \citep{srivastava2021sparse} algorithms. The OMP and SBL algorithms are employed in the SMV model, and applied on a subcarrier-by-subcarrier basis. At the end, the average of the estimated channel vectors is taken for the final channel estimation. The stopping parameter for the SBL, GSBL and modified TMSBL algorithms is set to $K_{\rm{max}}=100$. 

\subsection{Complexity Comparison}
\begin{figure*}[!hbt]
{
	\begin{minipage}[c][1\width]{
	   0.5\textwidth}
	   \centering
	   \includegraphics[width=1\textwidth]{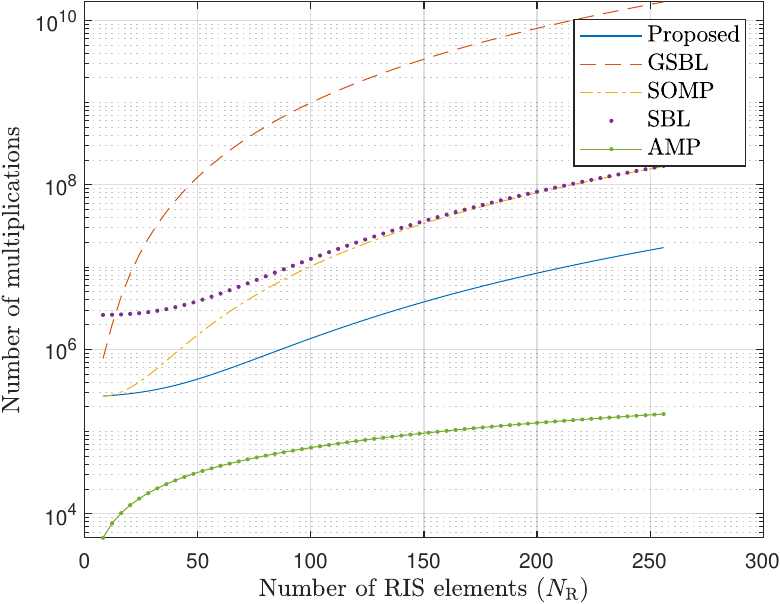}
          \caption{Computational complexity of different CS algorithms for $N=10$ and $J=64$, when different numbers of RIS element $N_{\text R}$ are employed.}
          \label{fig:complexity1}
	\end{minipage}}
 \hfill 	
{
	\begin{minipage}[c][1\width]{
	   0.5\textwidth}
	   \centering
	   \includegraphics[width=1\textwidth]{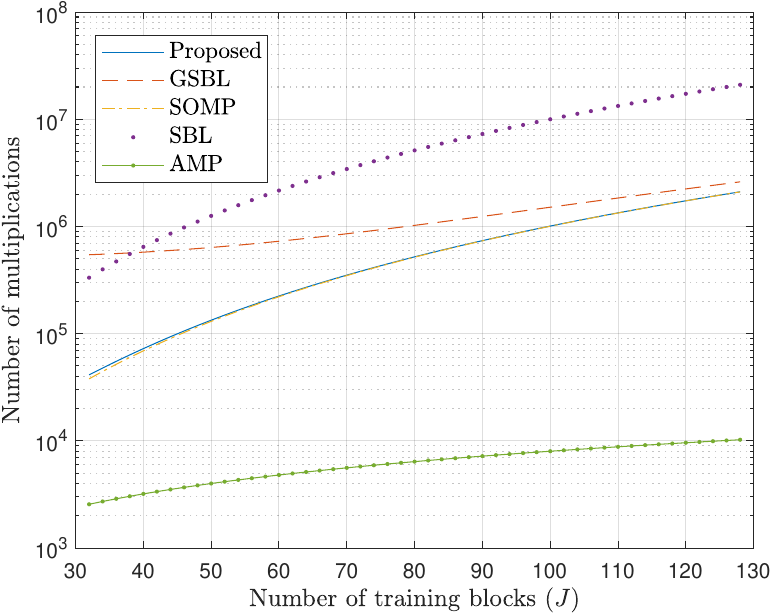}
          \caption{Computational complexity of different CS algorithms for $N=10$ and $N_{\text R}=8$, when different numbers of training blocks $J$ are employed.}
          \label{fig:complexity2}
	\end{minipage}}
 \hfill	
\end{figure*}

The complexity of the SBL algorithm is on the order of ${\textit{O}}(NN_{\rm{R}}^3+NJ^3)$ \citep{srivastava2021sparse}. The GSBL algorithm has a complexity order of ${\textit{O}}(N^3N_{\rm{R}}^3+J^3)$ based on \eqref{aposteriorievaluation.eq}. Explicitly, it is significantly higher than that of the SBL algorithm, as a result that the group sparsity increases the dimension of the involved dictionary matrix, whose inverse is required to compute. The modified TMSBL has the complexity order of ${\textit{O}}(N_{\rm{R}}^3+J^3+N_{\rm{R}}N^3)$ \citep{zhang2011sparse}, which can be significantly lower than that of the GSBL algorithm, due to the fact that it requires smaller dimension matrices in \eqref{sigmaapprox.eq} and \eqref{approxmu.eq}. For the SOMP algorithm, its complexity order is ${\textit{O}}(NN_{\rm{R}}^2+J^3)$ \citep{shahmansoori2017position}, which is lower than that of the SBL-based algorithms. For the AMP algorithm, the complexity order is $O(N_{\text{R}}NJ)$. Table~\ref{complexity.tab} compares the complexity of the considered algorithms, when setting $J=60$, $N=10$, and $N_{\rm{R}}=8$. Explicitly, the proposed algorithm and the DCS-SOMP algorithm have the lowest complexity among the considered algorithms. Furthermore, Fig.~\ref{fig:complexity1} and Fig.~\ref{fig:complexity2} demonstrate respectively the impact of the number of RIS elements $N_{\text{R}}$ and the number of training blocks $J$ on the computational complexity. Specifically from Fig.~\ref{fig:complexity1}, it can be found that the proposed and the AMP algorithm has a good performance in terms of computational complexity, especially when the number of RIS is large. As shown Fig.~\ref{fig:complexity2}, the SBL algorithm has the highest complexity, as it is operated on subcarrier-by-subcarrier basis, resulting in that the complexity of $O(NJ^3)$ contributed by matrix inverse is much higher than that of the rest algorithms. On the other hand, our proposed algorithm shows the lowest computational complexity, owing to that the size of the matrices needing inversion during the hyperparameter iterations is reduced. Hence, the proposed algorithm has a significantly lower complexity from the GSBL algorithm, especially when the system dimension is large.
\begin{table}[htbp]
\centering
\caption{Complexity comparison}
\begin{tabular}{|c|c|}
\hline
Algorithm      & Computational Complexity \\ \hline
DCS-SOMP       & ${\textit{O}}(216640)$                 \\ \hline
SBL            & ${\textit{O}}(2165120)$                \\ \hline
GSBL           & ${\textit{O}}(5336000)$                \\ \hline
Modified TMSBL & ${\textit{O}}(224512)$                 \\ \hline
AMP & ${\textit{O}}(4800)$                 \\ \hline
\end{tabular}
\label{complexity.tab}
\end{table}

\subsection{{Comparison of Channel Estimation Algorithms}}
\begin{figure}[!hbt]
	\centering
	\includegraphics[width=0.5\textwidth]{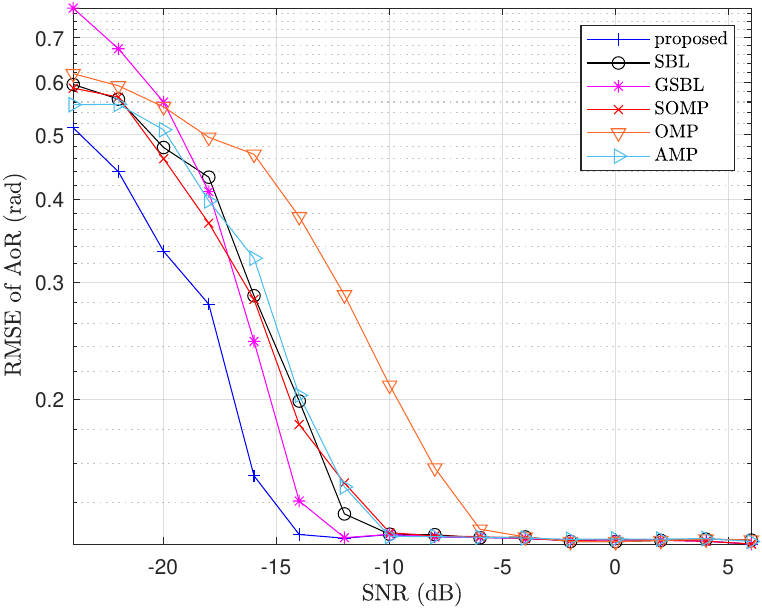} 
	\caption{RMSE performance of the estimated AoR for $N_{\text B}=N_{\text M}= N_{\text R}=8$ and $J=64$, when different channel estimation algorithms are employed.}
	\label{fig:AoRAlgorithm}
\end{figure}
Fig.~\ref{fig:AoRAlgorithm} compares the RMSE performance of the AoR estimation, when different channel estimation algorithms are employed. As shown in Fig.~\ref{fig:AoRAlgorithm}, the proposed modified TMSBL is capable of achieving more accurate estimation than the other considered algorithms. The AMP algorithm attains worse RMSE performance than the SBL-based algorithm, as the result that the AMP algorithm has experts the sensing matrix to satisfy RIP and follows near Gaussian distribution \citep{candes2008restricted}. However, it can be seen that for all the channel estimation algorithms, the RMSE floor appears, when SNR reaches a relatively high value. This is the result of the beamspace resolution, which unavoidably introduces quantization errors.

\begin{figure}[!hbt]
	\centering
	\includegraphics[width=0.5\textwidth]{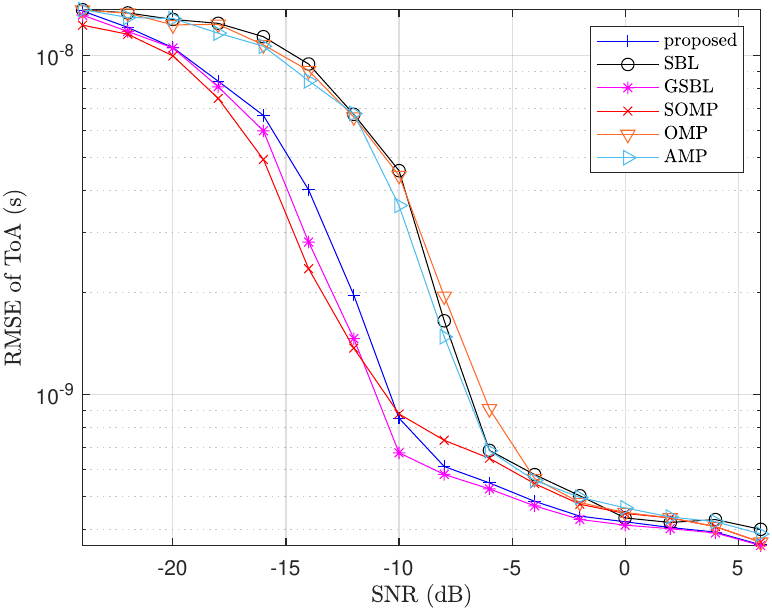} 
	\caption{RMSE of the estimated ToA for $N_{\text B}=N_{\text M}= N_{\text R}=8$ and $J=64$, when different channel estimation algorithms are employed.}
	\label{fig:ToAAlgorithm}
\end{figure}
The ToA estimation mainly relies on the channel estimation according to \eqref{timedelayRIS.eq}. As shown in Fig.~\ref{fig:ToAAlgorithm}, the GSBL algorithm attains the best RMSE performance compared to both the SOMP and the SMV-based schemes, such as, the OMP and SBL algorithms, when SNR is sufficiently high, typically higher than -12~dB. This is because the GSBL algorithm can make use of the group sparsity, while the SOMP, OMP and SBL algorithms do not utilize the group sparsity. The performance of the SOMP algorithm is inferior to the GSBL scheme, because it is sensitive to the dimension of the dictionary matrix ${\widehat{\bm{\Psi}}}$ and the employed stopping criterion. More explicitly, the SBL-based algorithm estimates the sparse coefficients from the posterior distribution, enabling to handle the uncertainty of signals and noise better than the DCS-SOMP algorithm, as mentioned in \citep{determe2016improving}. Consequently, the SBL-based algorithms are able to provide more accurate recovery in the ToA estimation. However, the computational complexity of the SOMP algorithm is lower than that of the SBL-based approaches. Therefore, there is a trade-off between RMSE performance and computational complexity. By contrast, when SNR is reasonably high, the proposed TMSBL algorithm is as effective as the GSBL algorithm, achieving the similar RMSE performance as the GSBL algorithm, but with a significantly lower computational complexity. Furthermore, as seen in Fig.~\ref{fig:ToAAlgorithm}, the AMP algorithm has worse RMSE performance in ToA estimation than the proposed and GSBL algorithms. This is because the AMP algorithm relies on the near Gaussian distributed sensing matrix. Otherwise, it performs poorer than the proposed and GSBL algorithm.

\begin{figure}[!hbt]
	\centering
	\includegraphics[width=0.5\textwidth]{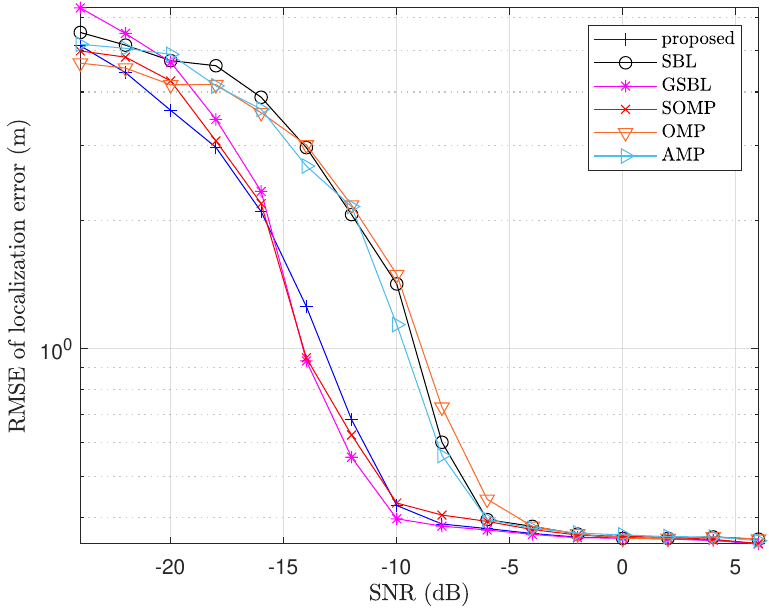} 
	\caption{RMSE performance of the estimated location of UE for $N_{\text B}=N_{\text M}= N_{\text R}=8$ and $J=64$, when different channel estimation algorithms are employed.}
	\label{fig:locAlgorithm}
\end{figure}
Based on the observations from Fig.~\ref{fig:AoRAlgorithm} and Fig.~\ref{fig:ToAAlgorithm}, the corresponding RMSE of position estimation is shown in Fig.~\ref{fig:locAlgorithm}. Again, the GSBL algorithm attains the best RMSE performance compared to both the SOMP and the SMV-based algorithms. The proposed TMSBL algorithm is also as efficient as the GSBL algorithm in terms of the RMSE performance, but outperforms the GSBL algorithm in terms of the computational complexity. Compared to the SOMP algorithm, the proposed TMSBL algorithm achieves better RMSE performance, when SNR$>$-10~dB, at the cost of the slightly increased complexity. For the AMP algorithm, due to its poor estimation of both channel gain and the corresponding ToA, as depicted in Fig.~\ref{fig:ToAAlgorithm}, it is unable to provide the positioning as accurate as the proposed algorithm, as shown in Fig.~\ref{fig:locAlgorithm}. Therefore, the AMP algorithm is in general not a good choice  for localization in the RIS-aided mmWave systems.

\subsection{Effect of Number of Training Blocks}
\begin{figure}[!hbt]
	\centering
	\includegraphics[width=0.5\textwidth]{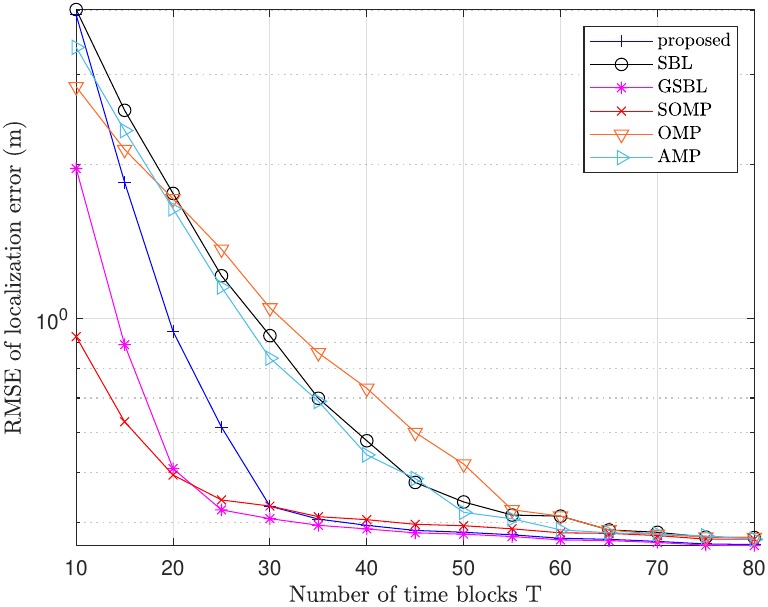} 
	\caption{RMSE performance of the estimated location of the UE versus the number of time blocks $J$ for $N_{\text B}=N_{\text M}= N_{\text R}=8$ and SNR$=$-5~dB, when different channel estimation algorithms are employed.}
	\label{fig:timeblock}
\end{figure}
Fig.~\ref{fig:timeblock} investigates the RMSE performance of the position estimation versus the number of training blocks $J$ at SNR=-5~dB. From the results, we can observe that when the number of training blocks $J<20$, the SOMP algorithm outperforms all the other algorithms considered. In detail, owing to the capability of the group sparsity awareness, the MMV model has superiority compared to the OMP and SBL algorithms. On the other hand, when $J$ is relatively small, the column in the dictionary matrix is less coherent. In this case, when $J$ increases, the GSBL algorithm outperforms the other the algorithms. 

\subsection{Effect of Number of RIS Elements (Resolution)}
\begin{figure*}[!hbt]
{
	\begin{minipage}[c][1\width]{
	   0.3\textwidth}
	   \centering
	   \includegraphics[width=1\textwidth]{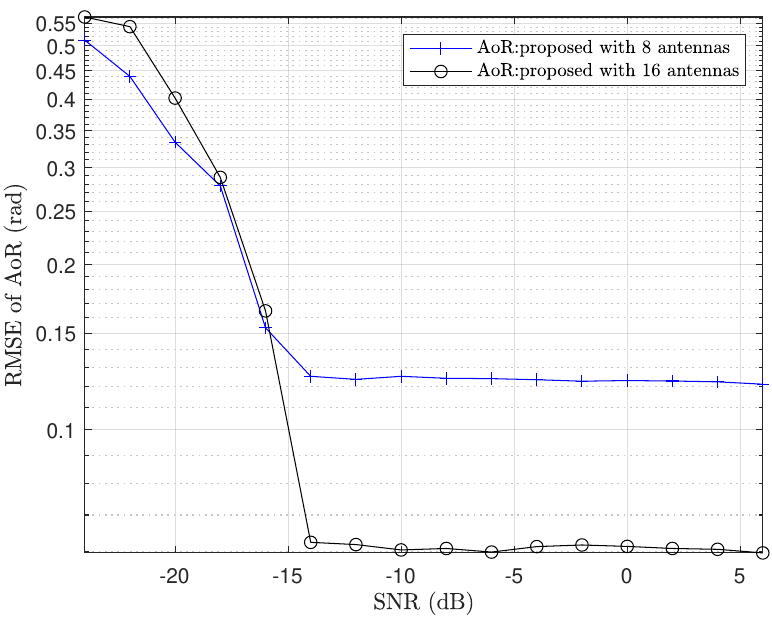}
	\end{minipage}}
 \hfill 	
{
	\begin{minipage}[c][1\width]{
	   0.3\textwidth}
	   \centering
	   \includegraphics[width=1\textwidth]{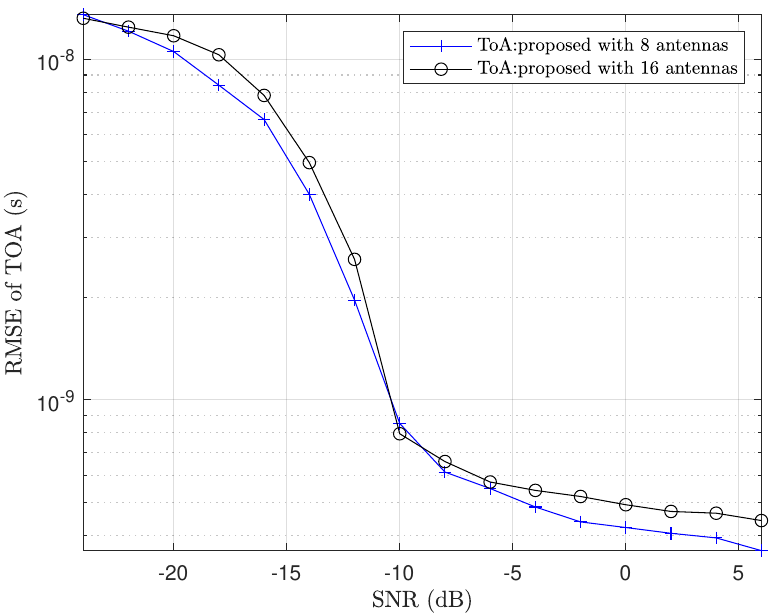}
	\end{minipage}}
 \hfill	
{
	\begin{minipage}[c][1\width]{
	   0.3\textwidth}
	   \centering
	   \includegraphics[width=1\textwidth]{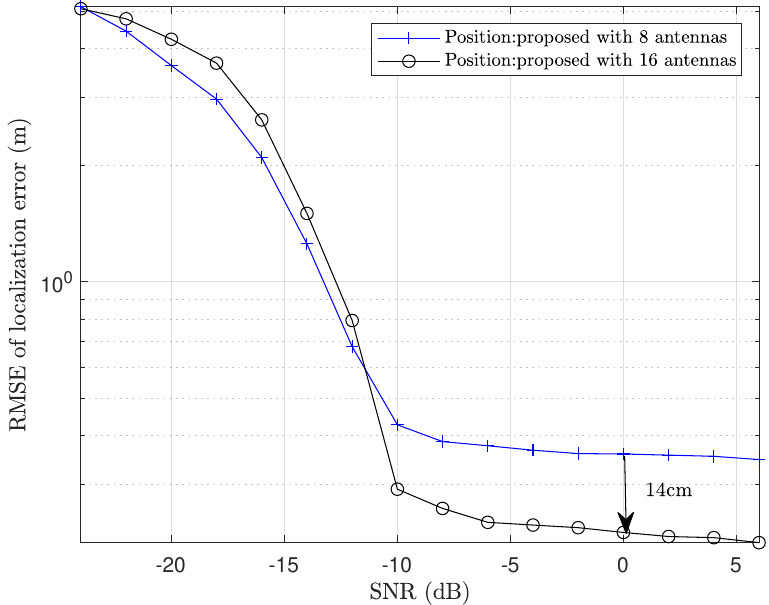}
	\end{minipage}}
\caption{RMSE performance of the estimated AoR, ToA, position of UE for $N=10$ and $J=64$, when $N_{\text B}=N_{\text M}= N_{\text R}=8$ and $N_{\text B}=N_{\text M}= N_{\text R}=16$ are employed.}
\label{fig:antennaimpact}
\end{figure*}
Fig.~\ref{fig:antennaimpact} shows the RMSE performance of localization, when RIS employs different number of reflection elements. As the total power of the RIS reflection is fixed, when increasing the number of RIS elements from 8 to 16, the error floor presents at the same SNR value. From the simulation results shown in Fig.~\ref{fig:antennaimpact}, we can be inferred that at relatively high SNR, such as SNR$=0$~dB, when the number of RIS elements increases, meaning that the beamspace resolution (size of the dictionary matrix) increases, which results in better performance in the estimation of AoR and ToA can be achieved. Specifically, the localization accuracy can be increased from about 0.36~meters to about 0.22~meters. However, when SNR is too low, such as SNR $=-10$~dB, employing more RIS elements results in worse performance in channel estimation and positioning. This is because, when the received signal is noise dominant, the estimator can only randomly select one possible value as the estimation result. In this case, when the beamspace resolution is higher owing to using more RIS elements, the probability of error is also higher, leading to worse positioning performance, as shown in Fig.~\ref{fig:antennaimpact}.

\subsection{Effect of Position of RIS}
\begin{figure}[!hbt]
	\centering
	\includegraphics[width=0.7\linewidth]{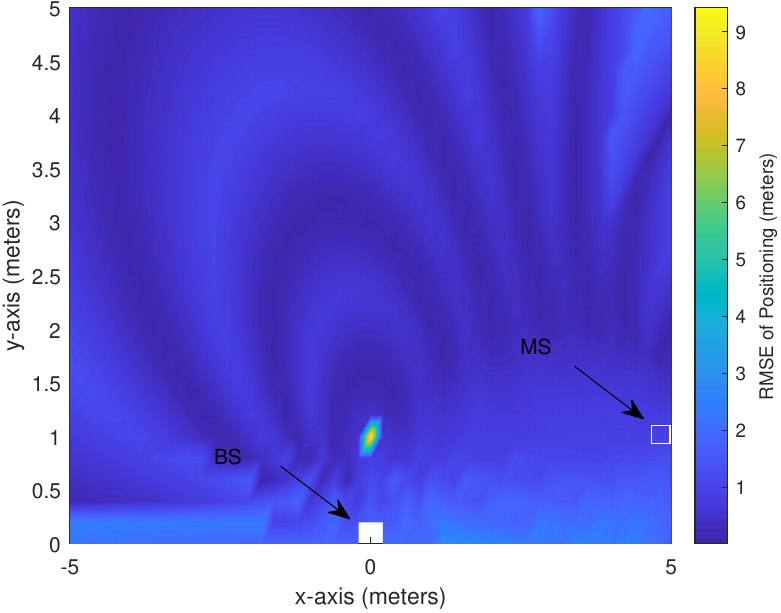} 
	\caption{RMSE performance with different RIS locations, when $N_{\rm{B}}=N_{\rm{R}}=N_{\rm{M}}=8$, where the location of BS is fixed at ${\bf{b}}=[0,0]^{\rm{T}}$, and the location of UE is at ${\bf{m}}=[5,1]^{\rm{T}}$.}
	\label{fig:RISplacement1}
\end{figure}
 To investigate the effect of the location of RIS, the location of BS is set at ${\bf{b}}=[0,0]^{\rm{T}}$, and the location of UE is set at ${\bf{m}}=[5,1]^{\rm{T}}$. We consider an indoor scenario, assuming a room of $5\times5$ and that the RIS can be deployed anywhere in the room. The localization is implemented by the proposed algorithm. From the results shown in Fig.~\ref{fig:RISplacement1}, it can be seen that the location of RIS has a big impact on the performance of the localization of UE. This is the result of the grid mismatch problem on the beamspace estimation. More explicitly, when the numbers of antennas/elements are set to $N_{\rm{B}}=N_{\rm{R}}=N_{\rm{M}}=8$, the beamspace angle range is divided into 8 portions, with each of $2\pi/8$. In this regard, the estimated angles mapped from the beamspace grid leads to the quantization error. As shown in Fig.~\ref{fig:RISplacement1}, the positioning accuracy presents a grid-like distribution.

\begin{figure}[!hbt]
	\centering
	\includegraphics[width=0.7\linewidth]{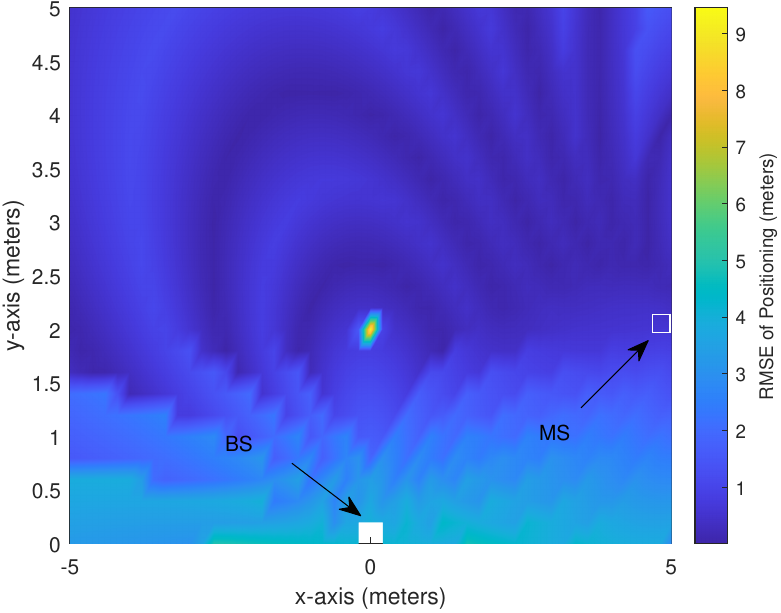} 
	\caption{RMSE performance with different RIS locations, when $N_{\rm{B}}=N_{\rm{R}}=N_{\rm{M}}=8$, the location of BS is at ${\bf{b}}=[0,0]^{\rm{T}}$, and the location of the UE is at ${\bf{m}}=[5,2]^{\rm{T}}$.}
	\label{fig:RISplacement2}
\end{figure}
Fig.~\ref{fig:RISplacement2} shows the RMSE performance of the localization of UE against the distribution of RIS, when BS is set at ${\bf{b}}=[0,0]^{\rm{T}}$, and UE is at ${\bf{m}}=[5,2]^{\rm{T}}$, and the numbers of antenna elements are set to $N_{\rm{B}}=N_{\rm{R}}=N_{\rm{M}}=8$. From Fig.~\ref{fig:RISplacement1} and Fig.~\ref{fig:RISplacement2}, we can find that when the RIS is placed with the same x-axis value as BS and the same y-axis value of UE, there appears an estimation error peak. This is because when the spatial frequency estimation is on the opposite side, it gives the same results, which leads to the ambiguity of localization by \eqref{GSBLAoR.eq}. Thus, in practice, the placement of RIS should avoid this kind of distribution, when the positioning relied services are provided. Moreover, RIS should be located at a position with a relatively high in y-axis value, as shown in Fig.~\ref{fig:RISplacement1}, in order to improve the localization coverage. 

\begin{figure}[!hbt]
	\centering
	\includegraphics[width=0.7\linewidth]{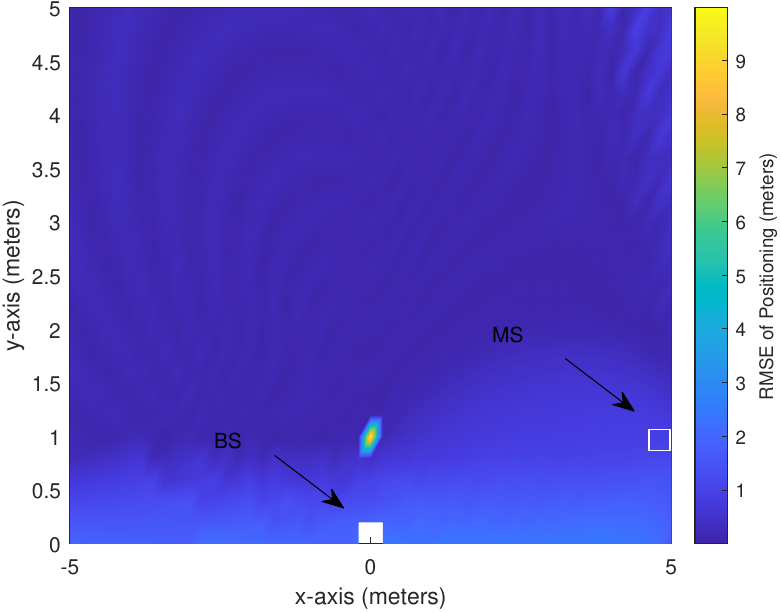} 
	\caption{RMSE performance with different RIS placement, when $N_{\rm{B}}=N_{\rm{R}}=N_{\rm{M}}=32$, and BS is at ${\bf{b}}=[0,0]^{\rm{T}}$, and UE is set at ${\bf{m}}=[5,1]^{\rm{T}}$.}
	\label{fig:RISplacement3}
\end{figure}
Finally, in Fig.~\ref{fig:RISplacement3}, we show the RMSE performance of the localization of UE with the distribution of RIS, when the numbers of antenna elements are $N_{\rm{B}}=N_{\rm{R}}=N_{\rm{M}}=32$. Similar to Fig.~\ref{fig:RISplacement2}, the location of BS is at ${\bf{b}}=[0,0]^{\rm{T}}$, and that of UE is at ${\bf{m}}=[5,1]^{\rm{T}}$. Compared to Fig.~\ref{fig:RISplacement1} corresponding to $N_{\rm{B}}=N_{\rm{R}}=N_{\rm{M}}=8$, the localization coverage shown in Fig.~\ref{fig:RISplacement3} is significantly improved, owing to the higher beamspace resolution, as shown in Fig.~\ref{fig:antennaimpact}. Furthermore, when comparing the two figures, the positioning accuracy in Fig.~\ref{fig:RISplacement3} presents a denser grid-like distribution.

\section{Conclusions}\label{conclusion.sec}

In this paper, the problem on the downlink mmWave channel estimation based localization was studied with a RIS-assisted system. Our localization process was divided into two stages. During the first stage, the conventional mmWave channel estimation is carried out to design the precoder and combiner used at BS and UE, respectively. Based on the initial design obtained from the first stage, in the second stage, a modified TMSBL algorithm was proposed to estimate the AoR at RIS and ToA at UE, which takes the advantage of the group sparsity existing in the multiple observation blocks. In performance studies, the impact from different numbers of antenna elements and time blocks, as well as from the placement of RIS was demonstrated. Our studies and results show that, with the aid of a RIS, the positioning accuracy can reach the centimeter level, even when the LoS link between BS and UE is blocked. However, when placing a RIS for positioning purpose, we should avoid placing it at the position that makes BS, RIS and UE form a rectangular with the RIS on the right angle. If this is the case, the beamspace angle mapping leads to the localization ambiguity, yielding large positioning error. The problem of localization ambiguity may be solved by employing two or more RISs, which provide extra benefits but also impose other challenges that expect further research.

\small
\bibliographystyle{IEEEtran}
\bibliography{RIS-Localization-10-23}

\end{document}